\documentclass[useAMS,usenatbib]{mn2e}

\usepackage{graphicx}
\usepackage{times}
\usepackage{natbib}

\newcommand{\enzo}{\it {\small ENZO}}

\begin{document}
 
\title[Efficiency of shock acceleration in clusters]{Constraining the efficiency of cosmic ray acceleration by cluster shocks}
\author[F. Vazza, M. Br\"{u}ggen, D. Wittor, C. Gheller,  D. Eckert, M. Stubbe]{F. Vazza$^{1}$\thanks{%
 E-mail: franco.vazza@hs.uni-hamburg.de}, M. Br\"{u}ggen$^{1}$, D. Wittor$^{1}$, C. Gheller$^{2}$, D. Eckert$^{3}$, M. Stubbe$^{1}$
\\
$^{1}$ Hamburger Sternwarte (Hamburg University), Gojenbergsweg 112, 20535 Hamburg, Germany \\
$^{2}$ ETHZ-CSCS, Via Trevano 131, CH-6900 Lugano, Switzerland\\
$^{3}$ Astronomy Department, University of Geneva 16, ch. d'Ecogia, CH-1290 Versoix Switzerland}

\date{Accepted ???. Received ???; in original form ???}
\maketitle

\begin{abstract}
We study the acceleration of cosmic rays by collisionless structure formation shocks with {\enzo} grid simulations. Data from the FERMI satellite enable the use of  galaxy clusters as a testbed for particle acceleration models. Based on advanced cosmological simulations that include different prescriptions for gas and cosmic rays physics, we use the predicted $\gamma$-ray emission to constrain the shock acceleration efficiency.
 We infer that the efficiency must be on average $\leq 10^{-3}$ for cosmic shocks, particularly for the  $\mathcal{M} \sim 2-5$ merger shocks that are mostly responsible for the thermalisation of the intracluster medium. These results emerge, both, from non-radiative and radiative runs including feedback from active galactic nuclei, as well as from zoomed resimulations of a cluster resembling MACSJ1752.0+0440. 
The limit on the acceleration efficiency we report is lower than what has been assumed in the literature so far. Combined with the information from radio emission in clusters, it appears that a revision of the present understanding of shock acceleration in the ICM is unavoidable.

\end{abstract}

\label{firstpage}

\section{Introduction}
\label{sec:intro}

The growth of cosmic structures naturally leads to the formation of powerful shock waves into the intergalactic and the 
intralcluster medium (IGM and ICM, respectively) (e.g. \citealt[][]{1972A&A....20..189S}, \citealt[][]{2013PhR...533...69C}). The Mach number of shocks within galaxy clusters at late epochs must be low because the 
mergers are between virialized halos \citep[][]{2003ApJ...583..695G}. On the contrary, stronger accretion
shocks  must be located at all epochs in the outer parts of large-scale structures, as they
mark the transition from infalling matter to the onset of the virialisation process  \citep[][]{mi00,ry03}. 
Cosmological simulations show
that the bulk of kinetic energy dissipation in the cosmological volume proceeds via shocks with Mach numbers $2 \leq \mathcal{M} \leq 3$ \citep[e.g.][]{va11comparison}. \\
Radio relics are steep-spectrum radio sources that are usually detected in the 
outer parts of galaxy clusters, $\sim 0.5-3$ Mpc from their centres and often found in clusters with a perturbed dynamical 
state \citet[e.g.][]{1998A&A...332..395E,hb07}. 
However, radio observations of radio relics \citep[][]{fe08,fe12} are biased towards merger shocks with $2 \leq \mathcal{M} \leq 5$, due to the larger weighting of radio-emitting electrons with a flatter spectral index \citep[][]{sk13,2014ApJ...785..133H}. 
Diffusive shock acceleration (DSA, \citealt[e.g.][]{2012JCAP...07..038C,kr13,2014ApJ...783...91C}) has been singled out as the most likely mechanism to
accelerate relativistic particles at cosmic collisionless shocks. However, the relative efficiency of this process for electrons and protons, and its dependence on the shock and plasma parameters are poorly constrained. 
Recently, hybrid simulations have shown that the diffusive shock acceleration of relativistic protons and the CR-driven amplification of magnetic fields in $5 \leq \mathcal{M} \leq 50$ shocks are efficient only for quasi-parallel configurations, $\theta \leq 45^{\circ}$ \citep[][]{ca14a}.

While the power-law emission spectra of radio-emitting electrons in radio relics are naturally explained by DSA \citep[e.g.][]{hb07,vw10},
recent modelling suggests that at least in some cases the measured acceleration of electrons is too large, and at odds with DSA \citep[e.g.][]{ka12,pinzke13}. 
The inclusion of re-accelerated particles can alleviate the tension in some cases, yet in order for the cosmic ray protons to be equally reaccelerated and become detectable in $\gamma$-rays, their injection efficiency must be much below the predictions from DSA \citep[][]{va14relics,va15relics,bj14}.

Our findings suggest that additional acceleration mechanisms might be responsible for channeling energy into the acceleration of radio-emitting electrons, while keeping the acceleration efficiency of protons rather low. Promising results in this direction have been recently obtained with Particle-In-Cell simulations 
\citep[][]{guo14a,guo14b}.


The release of relativistic protons into the ICM by cosmological shock waves, and the fact that they can be
stored there for longer than the Hubble time has been investigated in many works \citep[][]{1980ApJ...239L..93D,bbp97,bl99, pf07}. Interactions with the thermal ions of the ICM lead to diffuse hadronic $\gamma$-ray emission, in 
the range of what can already be tested by $\gamma$-ray observations  \citep[][]{aha09,alek10,ack10,arl12,alek12,fermi13,zand14}. To date, no diffuse $\gamma$-ray emission from the ICM has been detected by FERMI \citep[][]{ack10, fermi13}. With this information, one can 
set  upper limits on the amount of CR-protons in the ICM, of the order of
a few percent for the virial volume of clusters provided that the radial distributions are similar to that of thermal baryons \citep[][]{ack10,2013A&A...560A..64H,zand14,2014ApJ...795L..21G}. \\
In this work we analyze a large sample of simulated galaxy clusters and compare the expected level of hadronic $\gamma$-ray emission as a function of the assumed acceleration scenario to constraints from observations.

\section{Methods}

Our simulations are produced using original modifications for cosmic ray physics on top of the {\enzo} code \citep[][]{enzo14}, as presented in previous work  \citep[][]{scienzo,va13feedback,scienzo14}.
We assume the  WMAP 7-year cosmology \citep[][]{2011ApJS..192...18K} with
$\Omega_0 = 1.0$, $\Omega_{B} = 0.0455$, $\Omega_{DM} =
0.2265$, $\Omega_{\Lambda} = 0.728$, Hubble parameter $h = 0.702$, a normalisation for the primordial density power
spectrum $\sigma_{8} = 0.81$ and a spectral index of $n_s=0.961$ for the primordial spectrum of initial matter
fluctuations, starting the runs at $z_{\rm in}=30$.\\
Compared to the suite of simulations presented in \citet[][]{scienzo14} here we only analyse boxes of $300^3 ~\rm Mpc^3$ and $150^3 \rm Mpc^3$  simulated with $2048^3$ and $1024^3$ cells/DM particles, respectively. In the second case, several resimulations compare the effects of CRs and gas physics on the hadronic emission. The use of large grids with constant resolution yields a large sample of clusters with appropriate detail in describing the outer cluster regions, where most shocks occur \citep[][]{va11comparison}.  A study of resolution effects is provided in the Appendix. \\
Table 1 gives a schematic view of the runs used in this work.
All halos with  $\geq 10^{13} M_{\odot}$ are sampled by at least $\sim 10^4$ DM particles, ensuring a robust modelling of the innermost cluster dynamics.
At $z=0$ the CUR1 box contains $\approx 170$ halos with masses $M_{\rm v} \geq 10^{13} M_{\odot}$ and $\approx 400$ with masses $M_{\rm v} \geq 10^{14} M_{\odot}$ while the CUR2
boxes contains about $\sim 8$ fewer objects due to their smaller volume. 
Although our goal of properly resolving the injection of CRs in a large sample of galaxy cluster can only be presently achieved with large unigrid runs, in Sec.\ref{subsec:result_amr} we also discuss higher resolution simulations of a galaxy cluster where we used adaptive mesh refinement, in order to better compare with small-scales X-ray and radio features of an observed galaxy cluster.

\subsection{Cosmic ray physics}
\label{subsec:crs}  

The injection of CRs and their dynamical feedback on the evolution of baryons is incorporated 
into the PPM hydrodynamical method of {\enzo}, using a two-fluid approach \citep[][]{scienzo,scienzo14}. The efficiency of conversion between the shock kinetic energy flux and the CR-energy flux is set by the Mach number measured on-the-fly. 
The models of DSA  tested in these simulations follow from modelling of shock acceleration with 1-D diffusion-convection methods \citep[e.g.][]{kj07,kr13}. Following these results, we assume that a fraction $\eta$ of the kinetic energy flux across the shocks surface ($\Phi_{\rm kin}$) is converted into CR-energy: $\Phi_{\rm CR}=\eta(\mathcal{M}) \Phi_{\rm kin}$, where $\eta=\eta(\mathcal{M})$. Simulated shocks can both inject CRs and re-accelerate pre-existing CRs, and we model both processes  at run-time by appropriately rescaling the efficiency function as a function of the energy ratio between CR and gas upstream of each shock \citep[][]{scienzo14}. 
Away from shocks, the CR-fluid is advected using the fluxes from the PPM solver, assuming that CRs are frozen into the gas component by magnetic fields, which is realistic given the large timescale for CRs diffusing out of our $\geq 100 \rm kpc$ cells. The dynamical impact of CRs follows from the effective equation
of state of each cell, which comes from the energy-weighted ratio of the gas and CRs ultrarelativistic index, $4/3 \leq \Gamma_{\rm eff} \leq 5/3$. We also include Coulomb and hadronic losses for CRs based on \citep[][]{guo08}.

In this work, we tested the acceleration efficiency by \citet{kj07}, the (less efficient) acceleration scenario by \citet{kr13} and a more simplistic scenario with a fixed acceleration efficiency of $\eta=10^{-3}$ independent of the Mach number.  We have also tested the more recent studies of DSA with hybrid-simulations by \citet{ca14a}. In the absence of an analytical or interpolated prescription for the acceleration efficiency as a function of Mach number and shock obliquity (which would require an expensive exploration of parameters), we can roughly model the net DSA acceleration efficiency of this model by rescaling the efficiencies of the \citet{kr13} model for an appropriate constant. The acceleration efficiency measured by \citet{ca14a} for $\mathcal{M} \sim 5-10$ quasi-parallel shocks is $\approx 0.5$ of the efficiency in \citet{kr13} for the same Mach number range. If we assume the probability distribution of shock obliquities to be purely random ($P(\theta) \propto \sin(\theta)$), only $\approx 0.3$ of shocks have $\theta \leq 45^{\circ}$. Hence, by combining these two factors we can roughly mimic the effect of DSA in the model by \citet{ca14a} by rescaling the \citet{kr13} acceleration model by $f_{\rm CS}=0.15$. Obviously, this is a very crude assumption and we defer to future magneto-hydrodynamical simulations to study how the shock obliquity affects the injection of CRs (Wittor et al., in prep).

\subsection{Gas physics}
\label{subsec:gas}
In runs including radiative cooling of gas, we assume a constant composition of a fully ionized H-He plasma with a uniform metallicity of $Z=0.3 ~Z_{\odot}$. The APEC emission model \citep[e.g.][]{2001ApJ...556L..91S} has been adopted to compute the cooling function  of each cell at run-time as a function of temperature and gas density \citep[][]{enzo14}.
For the cold gas in the simulated volume, with temperatures $T \leq 10^4 \rm K$, we use the cooling curve of \citet{2011ApJ...731....6S}, which is derived from a complete set of of metals (up to an atomic number 30), obtained with the chemical network of the photo-ionization software {\it Cloudy} \citep[][]{1998PASP..110..761F}. The thermal effect of the UV re-ionization background  \citep[][]{hm96} is approximately modelled with a gas temperature floor within $4 \leq z \leq 7$ \citep{va10kp}. 
The implementation of feedback from AGN has been presented in \citet{va13feedback} and \citet{scienzo14}. The code releases bipolar thermal jets at high gas density peaks in the simulation, identified within all massive halos. Our feedback scheme starts at $z=4$ (``AGN'' model) or at a lower redshift, $z=1$ (``AGN-low'' model) and deposits $E_{\rm AGN} = 10^{59} \rm erg$ of thermal energy in the  two cells on opposite sides of the gas density peaks (tagged as  $n \geq n_{\rm AGN}=10^{-2} \rm cm^{-3}$ comoving), with a random direction along the coordinate axes of the grid. This method for AGN feedback is admittedly simplistic as it by-passes, both, the problem of monitoring the mass accretion rate onto the central black hole within each galaxy, and the complex small-scale physical processes which couple the energy from the black hole to the surrounding gas \citep[see discussion in][]{scienzo14}. However, the tests presented in this work (Sec.\ref{subsec:clusters}) show it can produce realistic profiles of clusters on $\geq 100$ kpc scales.

\begin{table}
\label{tab:sim}
\caption{List of the simulations used in this work. Column 1: size of the simulated volume. Column 2: number of grid cells. Column 3: spatial resolution. Column 4: physical implementations and run name. Column 5: diffusive shock acceleration efficieny of cosmic rays (KJ07=\citealt{kj07}, KR13=\citealt{kr13}, $10^{-3}$=constant efficiency, CS14=\citealt{ca14a}).}
\centering \tabcolsep 4pt 
\begin{tabular}{c|c|c|c|c}
 $L_{\rm box}$ & $N_{\rm grid}$ & $\Delta x$ & details & DSA-efficiency\\\hline
 $216\rm ~Mpc/h$ & $2048^3$ & $105 \rm ~kpc/h$ &CUR1 non-rad. & KR13 \\
 $108\rm ~Mpc/h$ & $1024^3$ & $105\rm ~kpc/h$ &CUR2 non-rad. & KR13 \\
 $108\rm ~Mpc/h$ & $1024^3$ & $105\rm ~kpc/h$ &CUR2 non-rad. & KJ07 \\
 $108\rm ~Mpc/h$ & $1024^3$ & $105\rm ~kpc/h$ &CUR2 cool.+AGN(high) & KR13 \\
  $108\rm ~Mpc/h$ & $1024^3$ & $105\rm ~kpc/h$ &CUR2 cool.+AGN(low) & KR13 \\
  $108\rm ~Mpc/h$ & $1024^3$ & $105\rm ~kpc/h$ &CUR2 non-rad. & $10^{-3}$\\
  $108\rm ~Mpc/h$ & $1024^3$ & $105\rm ~kpc/h$ &CUR2 cool.+AGN(low) & $10^{-3}$\\
   $108\rm ~Mpc/h$ & $1024^3$ & $105\rm ~kpc/h$ &CUR2 non-rad. & CS14\\
\end{tabular}
\end{table}

\subsection{Adaptive mesh refinement resimulations of MACSJ1752.0+0440}
\label{subsec:amr}

We have performed an additional adaptive mesh refinement (AMR) resimulation of one massive
galaxy cluster that shows a good X-ray and radio morphological similarity with the
cluster  MACSJ1752.0+0440, that we already studied in \citet[][]{2012MNRAS.426...40B}.
Similar to other clusters with double relics, this object represents an important testbed for particle acceleration in the ICM, because the
location of relics allows us to precisely constrain the timing and the parameters of the merger \citep[][]{va14relics,va15relics}.
This cluster has a total virial mass of $\approx 1.37 \cdot 10^{15} M_{\odot}$ at $z=0$, while this is 
 $\approx 0.65 \cdot 10^{15} \rm M_{\odot}$ at the epoch of a major merger at $z=0.3$, similar to 
what is inferred from the X-ray data of MACSJ1752.0+0440 at $z=0.366$. 
We produced multiple resimulations of this object by varying the modelling of CR injection and use the FERMI data to constrain the CR-acceleration scenario.
The initial conditions for this object are taken from set of nested grids sampling a volume of $264^3 \rm Mpc^3$, centred on the formation region of the cluster. AMR based on the  local gas/DM overdensity and on velocity jumps is used to increase the resolution in the cluster region additional $2^3$ times, down to a maximum resolution of $\approx 25 ~\rm kpc/h$ for a $\sim 80\%$ fraction of the ICM volume.  For a more details on the simulation procedure we refer the reader to \citet{va10kp}. 


\subsection{$\gamma$-ray observation of MACSJ1752.0+0440} 
\label{subsec:fermi_macs}

For this project we derived upper limits for the  $\gamma$-ray emission from the region of  MACSJ1752.0+0440 at $z=0.366$, analyzing the FERMI catalog. 

We analyzed the data of the Large Area Telescope on board FERMI at the position of MACS J1752 to set constraints on the $\pi^0$ emission arising from the interaction of cosmic-ray protons with the ambient intra-cluster medium. We follow the method presented in \citet{va15relics} to analyze the FERMI-LAT data. Namely, we use the \texttt{ScienceTools} v9r32p5 software package and the P7SOURCE\_V6 instrument response files. We construct a model for the expected $\gamma$-ray spectrum arising from $\pi^0$ decay by convolving the simulated cosmic-ray spectrum in the [0.2-300] GeV band with the proton-proton interaction cross section from \citet{2006PhRvD..74c4018K}. The observed spectrum is then fit with the model template and the significance of the signal over the background is estimated by computing the likelihood ratio between the best-fit model and the null hypothesis (i.e. no additional source), usually referred to as the test statistic (TS). This analysis yields a very mild improvement in the likelihood (TS=0.04), which indicates that no significant signal is detected at the position of MACS J1752. The 95\% upper limit to the source flux is $1.2\times10^{-9}$ photons/cm$^2$/s in the [0.2-300] GeV band. For more details on the data analysis procedure, we refer to \citet{2012A&A...547A.102H} and \citet{2013A&A...560A..64H}.

\begin{figure*}
    \includegraphics[width=0.99\textwidth]{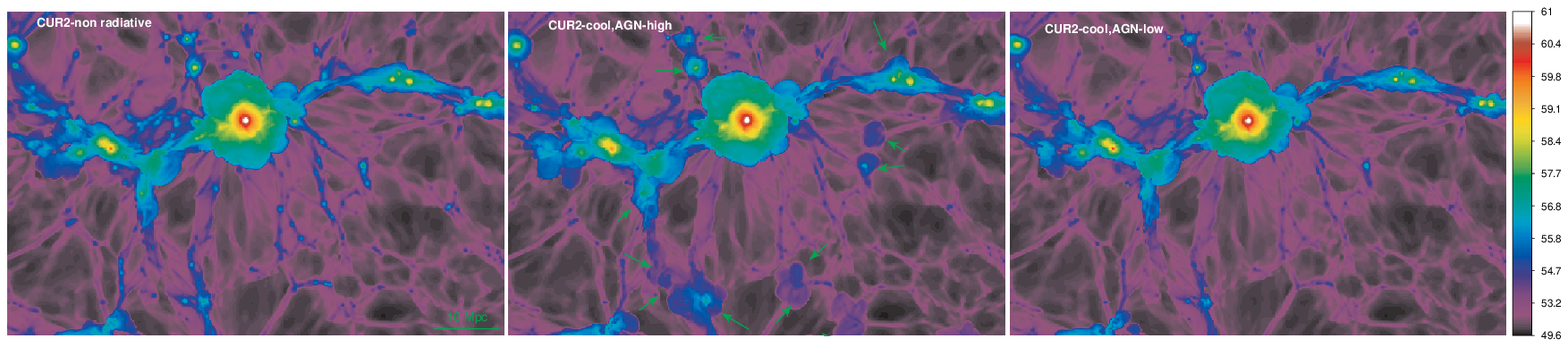}
     \includegraphics[width=0.99\textwidth]{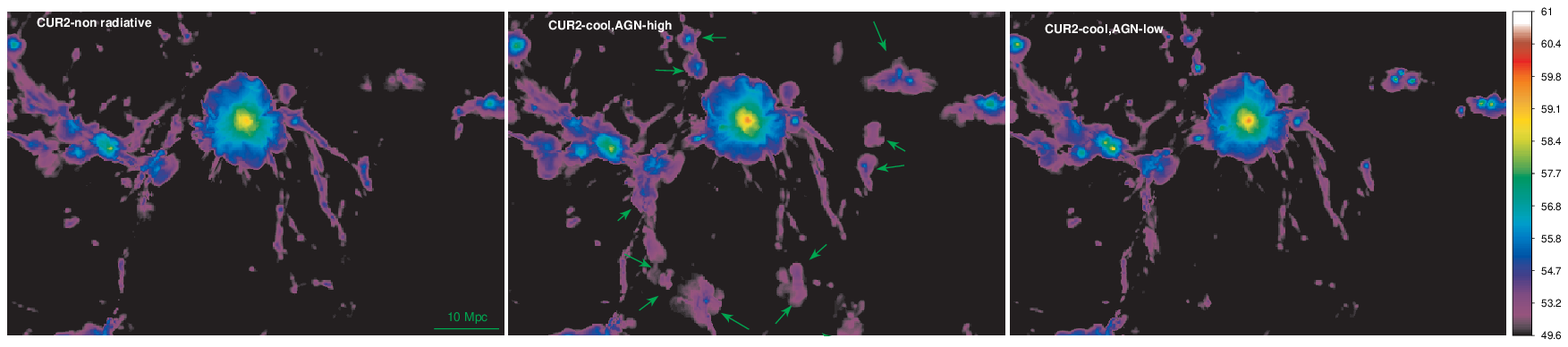}
     \caption{2-dimensional slice (with thickness $3$ Mpc) of the gas energy (top panels) and CR-energy (bottom panels) for a subvolume of the CUR2 volume at $z=0$, where we compare the non-radiative and the cooling plus AGN feedback runs. The color bar gives the energy per cell in units of $\log_{\rm 10} \rm [erg]$. To guide the eye, we indicate with green arrows the regions where the effect of AGN feedback is more prominent.}
  \label{fig:maps_first}
\end{figure*}

\begin{figure}
    \includegraphics[width=0.45\textwidth]{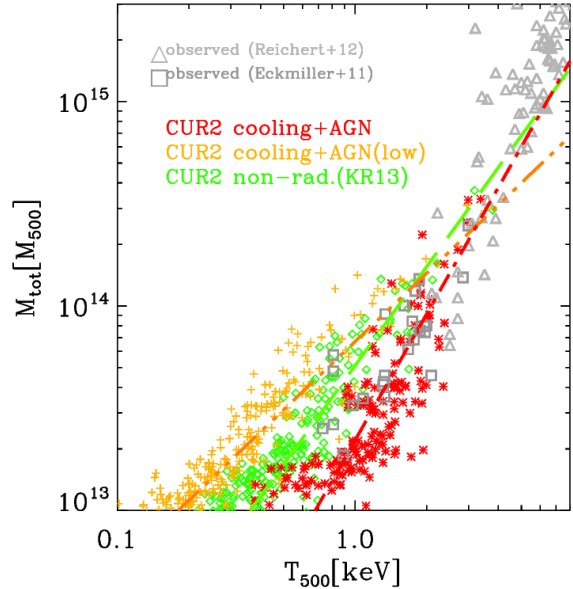}
    \caption{Mass-temperature scaling relation for the halos in the radiative and non-radiative runs of the CUR2 volume ($150^3 \rm Mpc^3$), computed inside
$R_{\rm 500}$ for each object at $z=0$. The additional lines show the best fit of the simulated data, while the two set of gray symbols are for real cluster observations using CHANDRA by \citet{2011A&A...535A.105E} and \citet{2011A&A...535A...4R}. To better compare with the simulated cluster and minimise the effect of cosmic evolution, we only consider observed cluster in the $0 \leq z \leq 0.2 $ redshift range.}
  \label{fig:scaling}
\end{figure}

\begin{figure*}
  \includegraphics[width=0.95\textwidth]{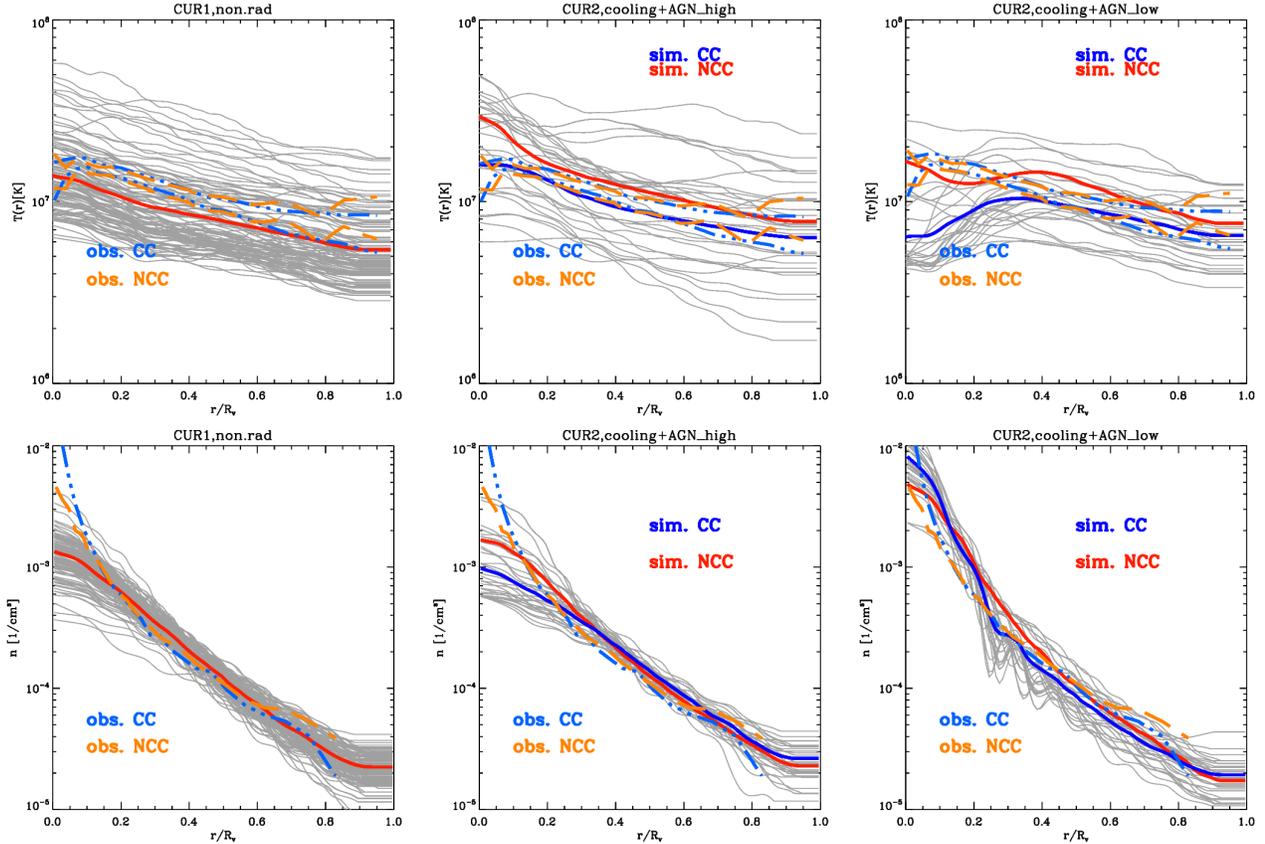}
  \caption{Radial profile of gas temperature and density for all simulated clusters with $M_{\rm vir} \geq 10^{14} M_{\odot}$ in the $300^3 \rm Mpc^3$ volume of the CUR1 run (non-radiative) and in the 
    $150^3 \rm Mpc^3$ volume of the CUR2 run (with cooling and two AGN feedback modes). The profiles of individual objects are shown in gray, while the $\pm \sigma$ around the mean profile of the sample are drawn with continuous lines (red lines for the NCC-like, blue line for the CC-like or the non-radiative clusters).  The additional lines shows the  $\pm \sigma$ around the mean profile of CC (dashed light blue) or the NCC (dot-dashed orange) from observations \citep[][]{2012A&A...541A..57E,2013A&A...550A.131P}.}  
 \label{fig:prof}
\end{figure*}

\section{Results}
\label{sec:results}

\subsection{Cluster properties}
\label{subsec:clusters}
The panels in Figure \ref{fig:maps_first} show the thermal and CR energy distribution in a subvolume of the CUR2 run with different prescriptions for gas physics and containing clusters of different masses. 
The impact of gas cooling and feedback is not very significant on the scale of the two massive objects in this image ($\sim 2 \times 10^{14} M_{\odot}$ and $\sim 4 \times 10^{14} M_{\odot}$, respectively), while it is more evident at the scale of galaxy groups, where AGN feedback promoted the expulsion of entropy and CR enriched gas outside of halos, as an effect of past powerful bursts. 
 Compared to the non-radiative case, the core of clusters/groups is always richer of CRs, due to additional injection following the shocks released by the thermal AGN feedback. The CR energy is everywhere smaller than the gas energy, with a ratio that goes from a few percent to a few tens of percent going from the centre to the outskirts of clusters. In the lower redshift AGN feedback case, the outer atmosphere of clusters is less extended in the other cases, due to the unbalanced compression by cooling.\\

In order to assess how realistic the thermal gas distribution in our clusters is, we compute
the mass-temperature relation for all identified halos in the volume, inside the reference overdensity of $\Delta=500$, and compare it to the X-ray observed scaling relations by \citet{2011A&A...535A...4R} and \citet{2011A&A...535A.105E}, see Fig.\ref{fig:scaling}. 


As expected, clusters in non-radiative runs closely follow the  self-similar scaling, $M \propto T^{3/2}$, while the cooling+AGN runs show significant departures from self-similarity. The run with lower redshift AGN feedback produces significant overcooling in small-size halos, which produces a {\it flattening} of the $(T,M)$ relation. In these runs the AGN feedback is just sufficient to quench the cooling flow, but most
clusters below  $M_{\rm 500} \leq 10^{14} M_{\odot}$ are too cold compared to observations. 
On the other hand, the run with early AGN feedback produce a scaling relation which is in better agreement with observations, with hints of a steepening for  $M_{\rm 500} \leq 10^{14} M_{\odot}$. No objects with a central temperature below $\sim 0.5 ~\rm keV$ are formed in this case. The scatter in temperature is also increased due to the
intermittent AGN activity. \\
We conclude that our fiducial AGN model is suitable to produce clusters with a realistic mass-temperature (and hence gas energy) relation, and can therefore represent a robust baseline model to test the outcome of different
CR-acceleration models against FERMI data. 
However, the lack of spatial resolution and physics
 at the scale of galaxies in these runs makes it impossible to  model star formation and star feedback (both energetic and chemical), and to properly compare the outcome of this against observed relations \citep[e.g.][for recent reviews]{2013MNRAS.428.1643P,2015ApJ...813L..17R,2015arXiv150904289H}. In the Appendix (Sec.\ref{appendix2}), we also show that the impact of CRs on the X-ray scaling relations is negligible in both tested acceleration models. \\

In Fig.\ref{fig:prof} we show the average radial profiles of gas temperature and density for our clusters, for the CUR1 and the CUR2 run with cooling and high redshift AGN feedback.
The results are compared to the observed mean profiles of gas density and temperature \citep[][]{2012A&A...541A..57E} and pressure \citep[][]{2013A&A...550A.131P} derived from X-ray and SZ observations of nearby clusters \citep[][]{2012A&A...541A..57E,2013A&A...550A.131P}.
Observations reported the consistent detection of a bimodal gas distribution in clusters having a cool-core (CC) or a without it (NCC), which shows up prominently as a difference of the innermost density, temperature and entropy profiles \citep[e.g.][]{2001ApJ...551..153D,cav09,2010A&A...513A..37H} as well as a smaller large-scale radius difference in density \citep[][]{2012A&A...541A..57E}.

We therefore split our cluster samples into CC and NCC classes based on the central temperature gradient observed
in each object at $z=0$. This is one of the several possible working definitions proposed in the literature \citep[e.g.][]{2010A&A...513A..37H}, which work well for the coarse resolution we have for
the central regions of the the lowest mass systems in the sample.  The gradient is defined as $\Delta T=T(r+\Delta r)-T(r)$ based on the spherical mass-weighted temperature profile, $T(r)$, and we consider a cluster CC-like if $\Delta T \geq 0$ in the first radial bin, or NCC-like otherwise.\\
All objects with $M_{\rm vir} \geq 5 \cdot 10^{13} M_{\odot}$ of the non-radiative CUR1 dataset are considered as NCC according to this criterion.
In the CUR2 sample with cooling and low redshift feedback case, we find a NCC/CC ratio close to $0.5$, yet 
all our objects are characterized by a too large central density, similar to or exceeding the one of CC systems. On the other hand, with the adoption of cooling and efficient feedback the ratio becomes NCC/CC $\sim 0.32$, i.e. quite close to observations \citep[i.e. $\approx 0.39$,][]{2010A&A...513A..37H}.\\
While the gas density profiles of clusters with different masses can be averaged, in averaging the temperature profiles we normalized each profile at $R_{\rm 500}$, based on the self-similar $T_{\rm 500} \propto M_{\rm 500}^3/2$ relation \citep[e.g.][]{2013A&A...551A..22E}.
The large-scale trends of thermodynamical
quantities are reproduced reasonably well by our runs. The cluster population in the CUR1 non-radiative run present an overall good match of the observed profiles in the
range $0.1 \leq R/R_{\rm vir} \leq 0.9$, but cannot reproduce the CC/NCC bimodality. The cluster population of the CUR2 run with cooling and early feedback does a similarly good job and shows hints of the CC/NCC bimodality. However, in this case the innermost density/pressure profile is underestimated compared to observations, due to the fact that the average cluster mass in this smaller volume is smaller and an increasing ratio of clusters has a core which is relatively poorly resolved compared to the CUR1 run, where many larger clusters are formed. We stress that while this effect may cause an underestimate of the hadronic $\gamma$-ray emission from the innermost regions (as this scales as $\propto n^2$) the effect is overall not large because typically only $\leq 10 \%$ of the $\gamma$-emission is produced within $\leq 0.1 R_{\rm vir}$ (see Sec.A2).
On the other hand, the cluster population in the CUR2 run with low redshift feedback produces typically too cold and dense CC clusters compared to observations, as an effect of overcooling. \\
In summary, the comparison with observations suggests that the clusters in the  non-radiative CUR1 run as well as in the cooling+AGN CUR2 run can be further used to study cosmic ray acceleration.  These cluster populations are representative enough of the global ICM properties for a wide range of masses/temperature to allow a comparison with FERMI observations in a similar range of masses. In particular, while clusters in the CUR1 can best represent the high-mass end of the observed distribution ($\geq 5 \cdot 10^{14} M_{\odot}$), the clusters in the CUR2 run can be used to better study the CC and NCC populations in the lower mass end. 
%

\begin{figure}
    \includegraphics[width=0.495\textwidth]{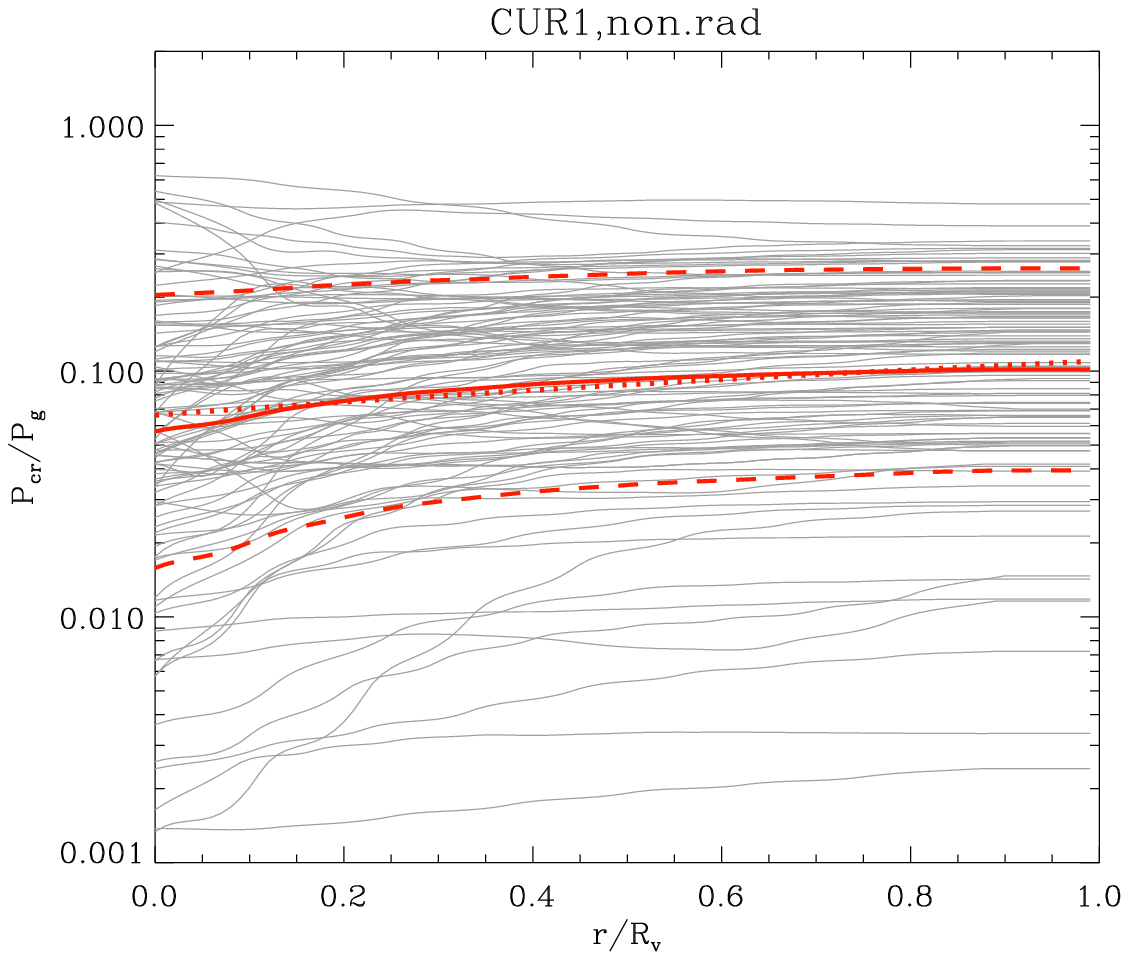}
\includegraphics[width=0.495\textwidth]{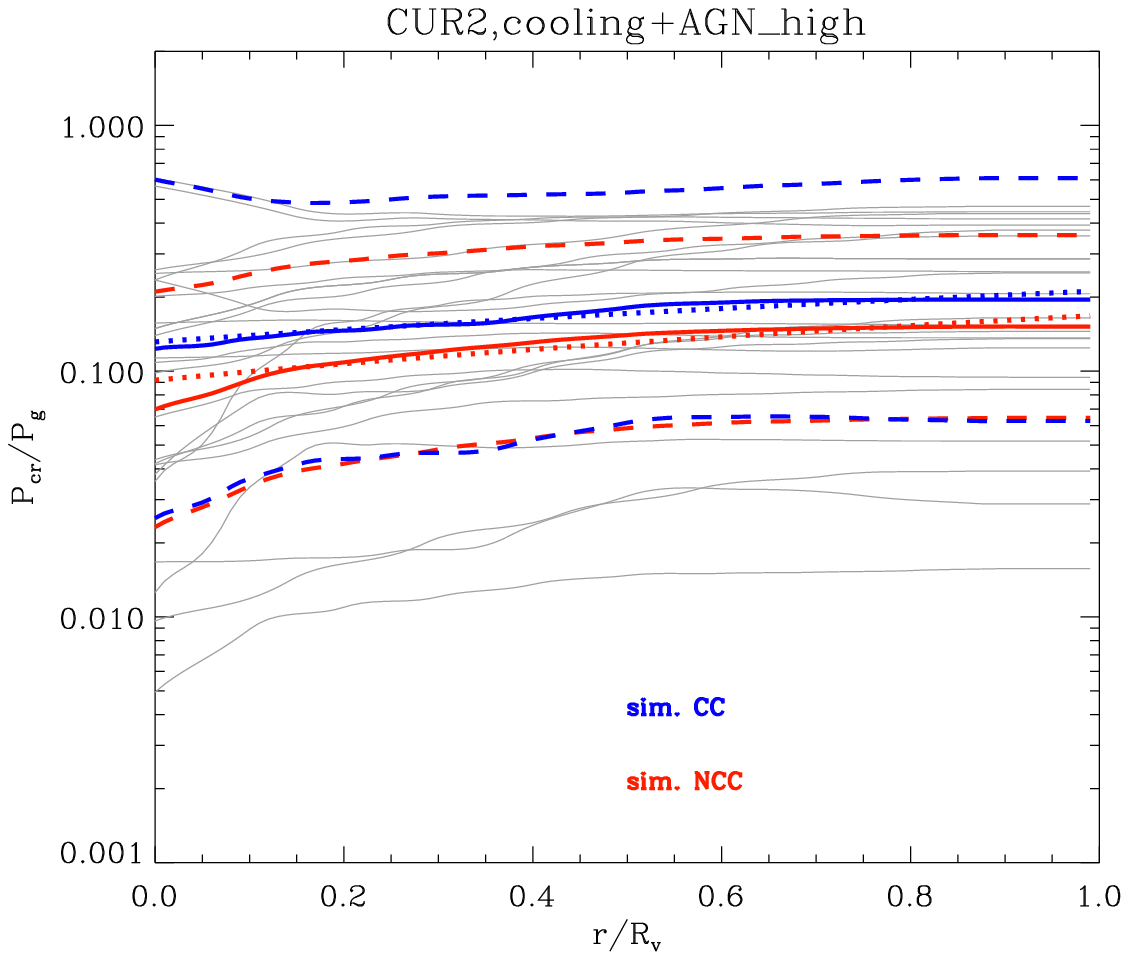}
 \caption{Average radial profile of the CR to gas pressure ratio for all simulated clusters in the CUR1 and CUR2 run with cooling and feedback, in all cases for the \citet{kr13} model of CR acceleration. The gray lines give the profiles of individual clusters while the coloured lines give the mean and the $\pm \sigma$ dispersion. The additional thin coloured line dotted lines give the best fit for the average profiles, with parameters given in Tab.\ref{tab:fit}.}  
 \label{fig:cr_profile}
\end{figure}

\begin{figure}
    \includegraphics[width=0.495\textwidth]{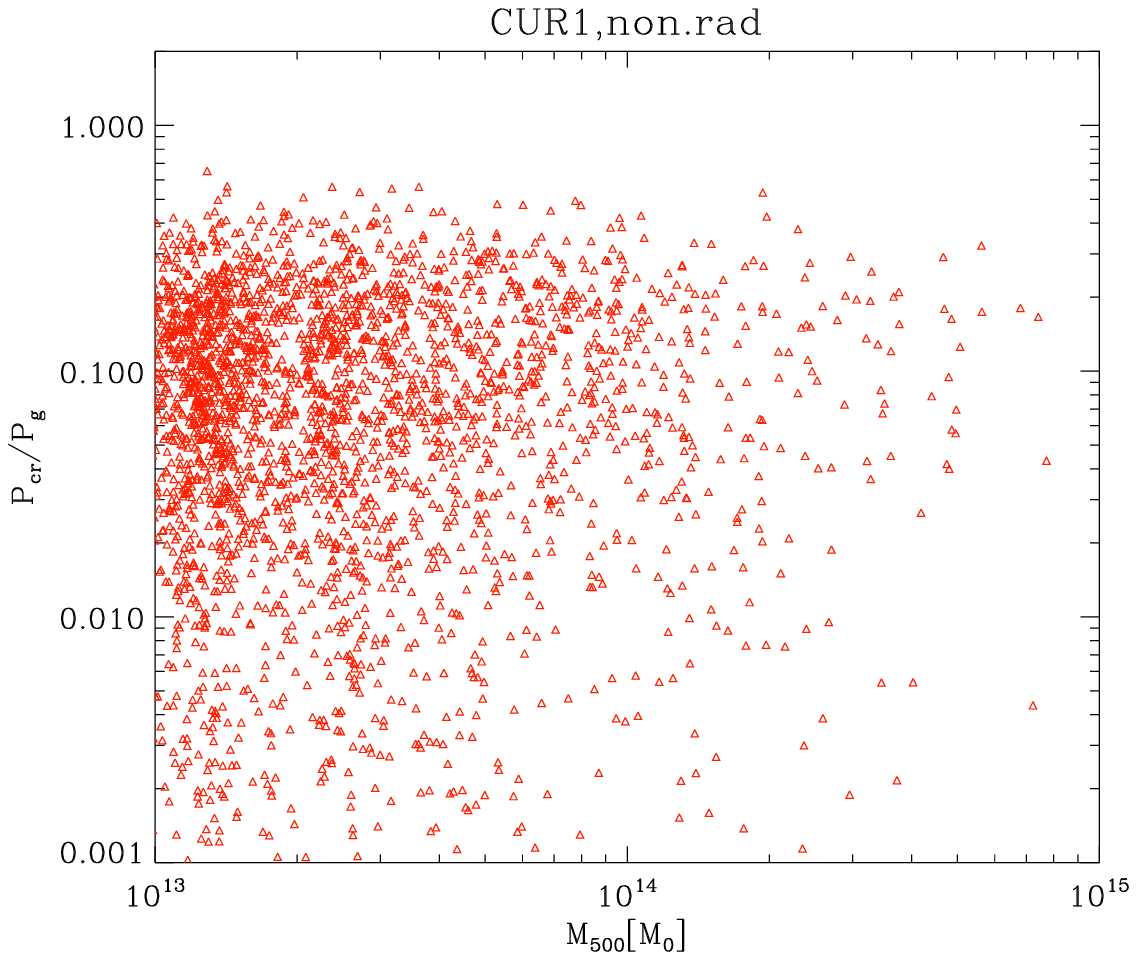}
    \includegraphics[width=0.495\textwidth]{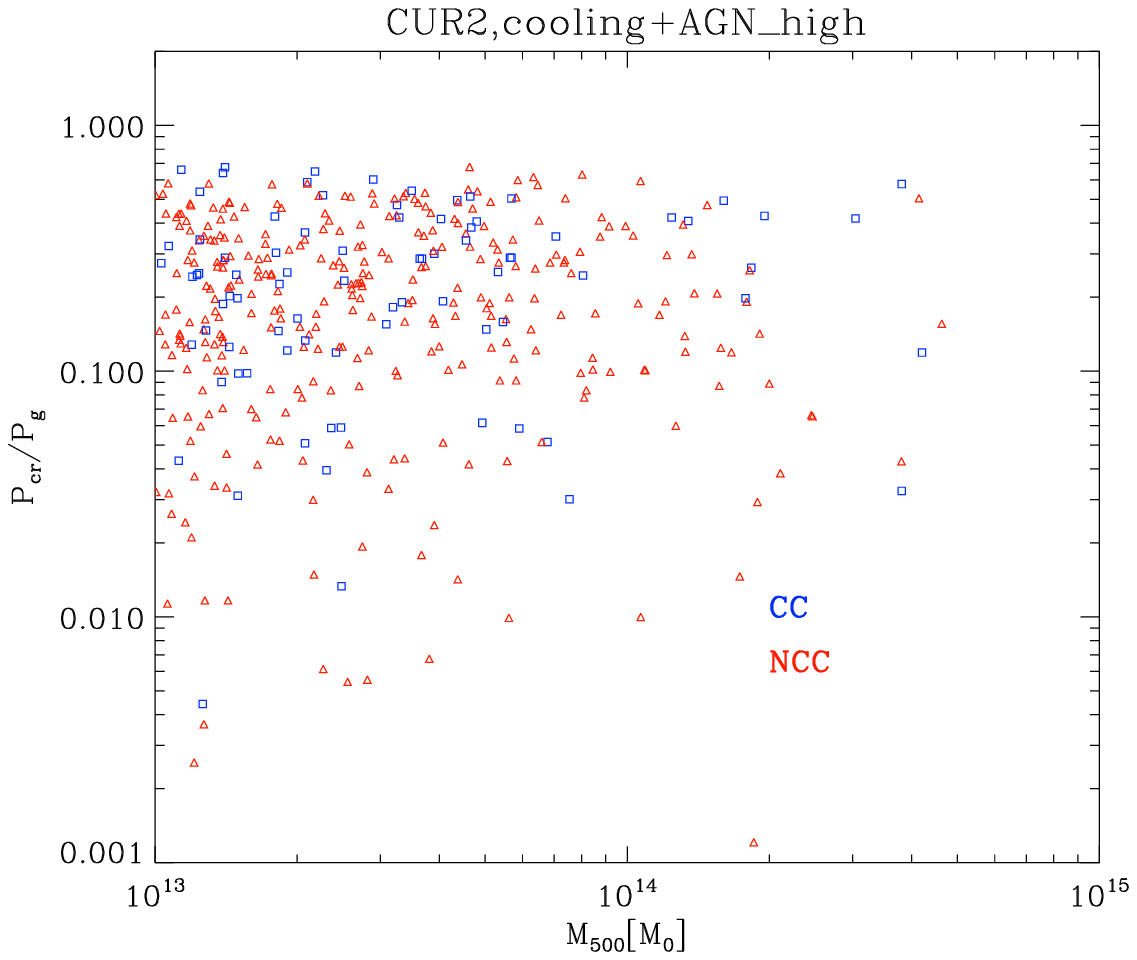}
    \caption{Average enclosed pressure ratio of CR and gas for simulated clusters at $z=0$. The top panel shows the distribution for the CUR1 run, the bottom panel shows the distribution for the CC- and NCC-like clusters in the CUR2 run with cooling and high redshift AGN feedback.}
  \label{fig:cr_scaling}
\end{figure}

\subsection{Cosmic ray properties}
\label{subsec:cr}

\begin{figure}
    \includegraphics[width=0.45\textwidth]{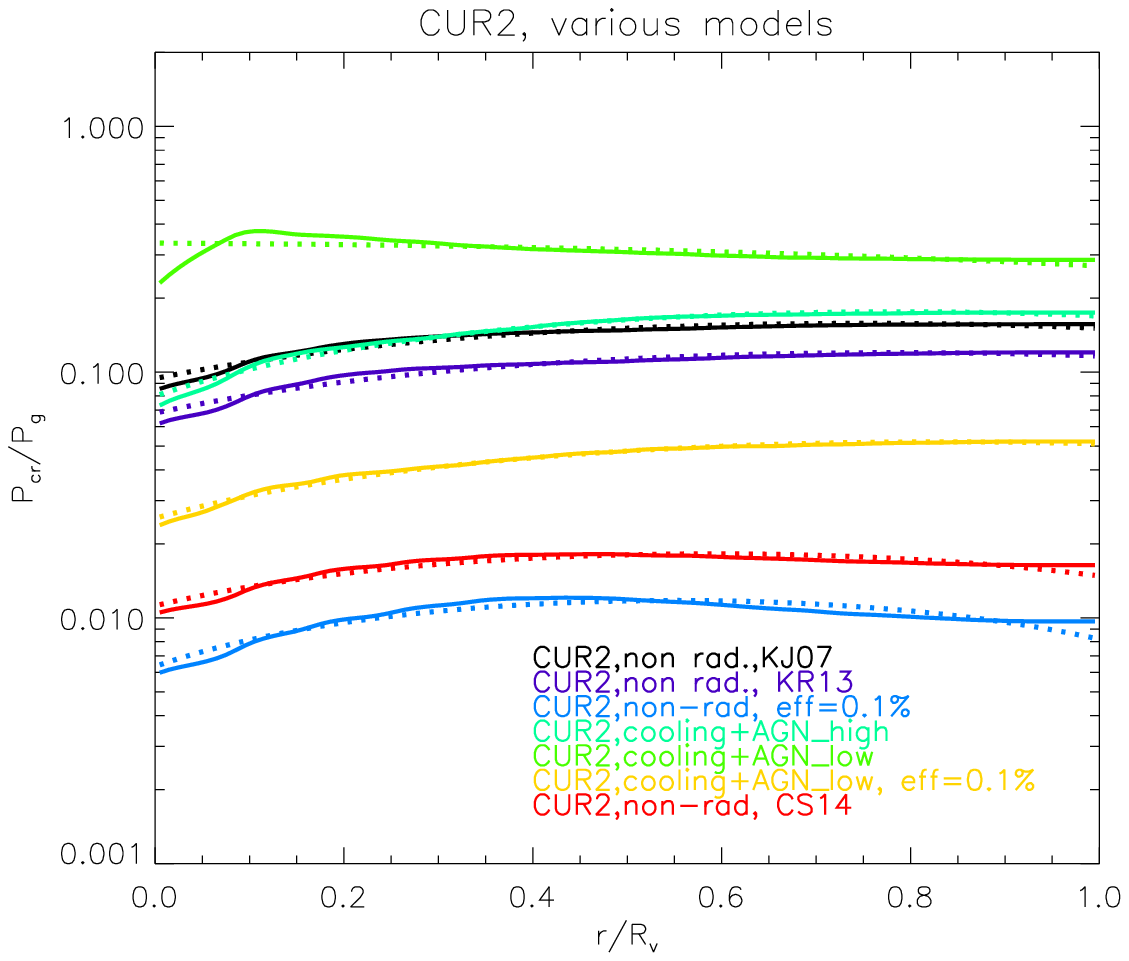}
    \caption{Average profile of $X(R)$ for all $M \geq 10^{14} M_{\odot}$ clusters in the CUR2 run, comparing different physical prescriptions for CRs and baryons. The dotted lines give the
    best fit relation for each model, with parameters given in Tab.2, whereas the dashed lines give the $\pm 1 \sigma$ standard deviations on the average profiles.}
  \label{fig:cr_prof2}
\end{figure}

\label{subsubsec:cr_profiles}

We extracted the spatial distribution of CR-energy for each simulated cluster at $z=0$, and estimate the relative pressure ratio of CRs compared to the thermal gas pressure. This is computed by integrating the total pressure of CRs within increasing radii, and dividing this by the total gas pressure within the same radius, $X(R)=P_{\rm cr}(<R)/P_{\rm g}(<R)$.
The pressure of CRs is  $P_{\rm  CR} = (\Gamma_{\rm CR}-1) E_{\rm CR}$, where $E_{\rm CR}$ is the primitive variable simulated with our two-fluid method (Sec. \ref{subsec:crs}). We consider the ultra-relativistic equation of state for CRs,  $\Gamma_{\rm CR}=4/3$, and therefore the pressure ratio can be written as  $X=P_{\rm CR}/P_{\rm g}=[(\Gamma_{\rm CR}-1)/(\Gamma-1)] E_{\rm CR}/E_{\rm g}=0.5 E_{\rm CR}/E_{\rm g}$.

The panels of Figure \ref{fig:cr_profile} give the profiles of $X(R)$ of the simulated clusters at $z=0$, Here we limit to the CUR1 run and to the CUR2 radiative run with high redshift AGN feedback, where we further split the sample into CC-like and NCC-like objects.
The average profiles are always very flat up to the virial radius, and are well fitted by a 2nd order polynomial,  
\begin{equation}
X(R) = X_0 + \alpha_1 \frac{R}{R_{\rm vir}} + \alpha_2 (\frac{R}{R_{\rm vir}})^2, 
\end{equation}
with best fit parameters given in Table 2 for all runs.

On average, the pressure ratio in the core of NCC-like clusters is $\sim 5-7 \%$, while this is $\sim 10 \%$ in CC-like clusters, while in all cases this ratio settles to $\sim 10-15 \%$ at $R_{\rm vir}$. However, the scatter in the cluster samples is rather large, as shown in Fig.\ref{fig:cr_scaling} where we show the relation between the cluster mass within $R_{\rm 500}$ and the total enclosed CR-to-gas pressure.  The data does not show a tight correlation of $X$ with cluster mass or cluster average temperature (not shown), and for every
mass bin variations of the ratio in the range $X \sim 10^{-3} - 0.3$ can be found. This is at variance with the earlier results by \citep[][]{2010MNRAS.409..449P}, based on a different numerical methods, where a tight decreasing correlation of $X$ with the host cluster mass is found. The extreme flatness of our profiles is also not in agreement with earlier SPH results \citep[][]{pf07}, while it is much more 
similar to more recent SPH simulations \citep[][]{2010MNRAS.409..449P}, indicating that numerical details play a significant role in the spatial distribution of CRs.  
We give our interpretation for this comparison in Sec.~\ref{sec:discussion}.

Figure \ref{fig:cr_prof2} shows the average profile of CR-to-gas pressure ratio for all $M \geq 10^{14} M_{\odot}$ clusters at $z=0$, where we compare the outcome of the different
models for gas and CR physics. The best-fit parameters in this case are given in the lower half of Table \ref{tab:fit}. Basically all models predict the same flat shape, with a minimum of $X(R)$ in the 
cluster core. In this set of models we also show the case of a {\it fixed} (re)acceleration efficiency of $\eta=10^{-3}$ and of the approximated acceleration model we derive from \cite{ca14a}. In the $\eta=10^{-3}$ model  both radiative and non-radiative runs also predict a very flat profile outside of clusters core, with a minimum of  $X \approx 0.6 \%$ in the non-radiative run and $X \approx 2 \%$ in the cooling and AGN case. The fact that the same fixed acceleration efficiency gives a $\sim 5$ times increased CR-to-gas pressure ratio stresses how much baryon physics can affect the modelling of CRs in the intracluster medium, and the 
quantitative interpretation of $\gamma$-ray data (see Sec.\ref{subsec:result_cosmo}). The  profiles obtained with the \cite{ca14a} model have a similar shape, with a $\sim 2$ higher normalisation.

\begin{table}
\label{tab:fit}
\caption{Best fit parameters for the $X,R$ relation for clusters in the CUR1 and CUR2 run, assuming $X(R) = X_0 + \alpha_1 (R/R_{\rm vir}) + \alpha_2 (R/R_{\rm vir})^2$. In the upper half of the table,  we give the best fit for the CC and NCC-like clusters separately, in radiative runs. The lower half of the table gives the best fit parameters limited to all $M\geq 10^{14} M_{\odot}$ clusters in the CUR2 runs.}
\centering \tabcolsep 5pt
\begin{tabular}{c|c|c|c}
 run & $X_0$ & $\alpha_1$ & $\alpha_2$ \\ \hline
 CUR1 & 0.056 & 0.099 & -0.055  \\
 CUR2, cool+AGN high (CC) & 0.117 & 0.162 & -0.082\\
 CUR2, cool+AGN high (NCC)  & 0.072 & 0.198 & -0.116     \\
 CUR2, cool+AGN low (CC) & 0.497 & -0.373 & -0.192\\
 CUR2, cool+AGN low (NCC) &  0.367 & -0.030 & -0.044\\\hline
  CUR2, non-rad, KJ07  & 0.094 & 0.170 & -0.113  \\
 CUR2, non-rad, KR13  & 0.068 & 0.132 & -0.084  \\
 CUR2, non-rad, $10^{-3}$ & 0.006  & 0.019 & -0.017 \\
 CUR2, non-rad, CS14 & 0.011 & 0.023 & -0.020 \\
 CUR2, cool+AGN high, KR13 & 0.079 & 0.244 & -0.155\\
 CUR2, cool+AGN low, KR13  & 0.334 & -0.012 & -0.052     \\
 CUR2, cool+AGN high, $10^{-3}$ &  0.025 & 0.067 & -0.038\\
\end{tabular}
\end{table}


\begin{figure}
    \includegraphics[width=0.42\textwidth,height=0.40\textwidth]{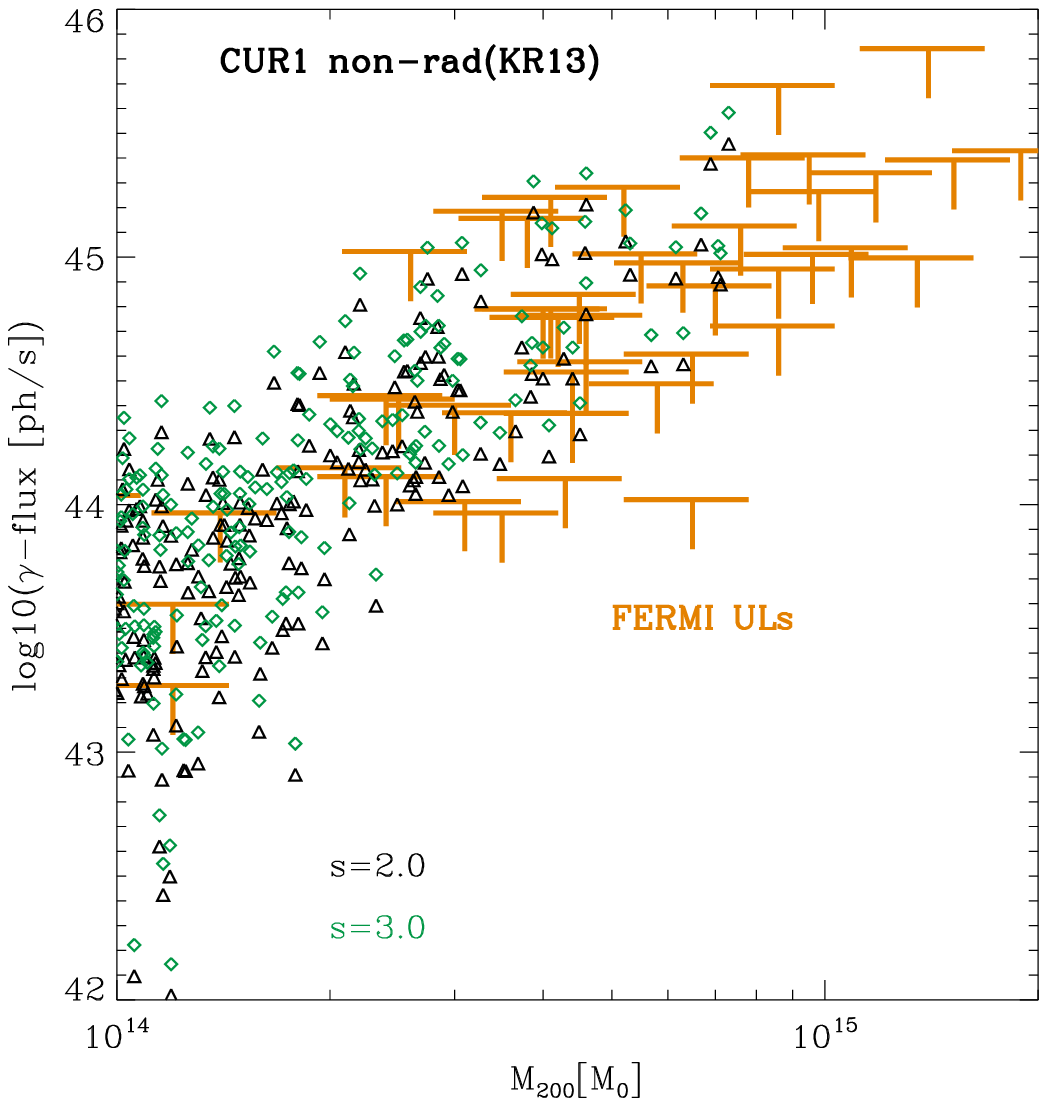}
    \includegraphics[width=0.42\textwidth,height=0.40\textwidth]{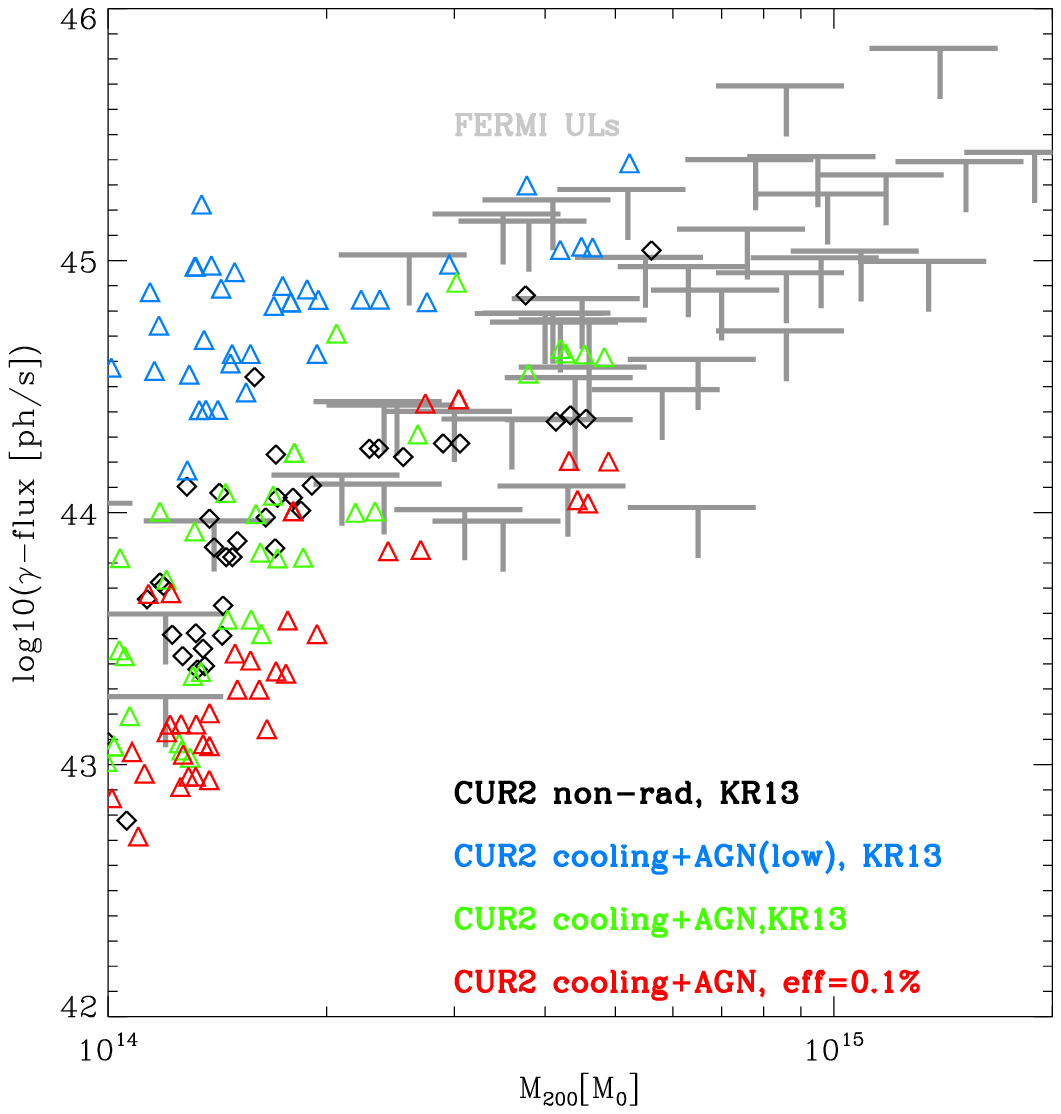}
    \includegraphics[width=0.42\textwidth,height=0.40\textwidth]{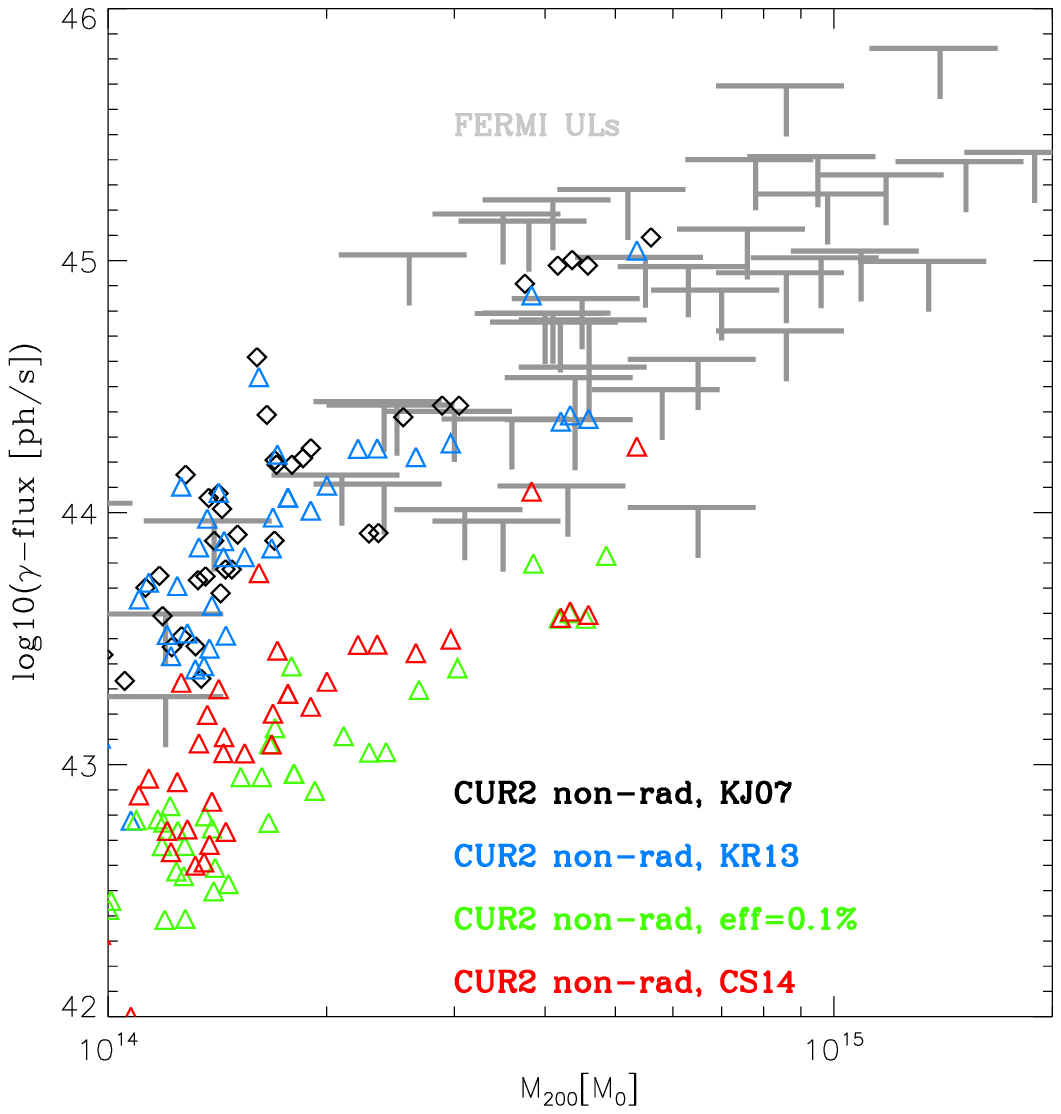}
        \caption{Hadronic emission for our simulated clusters at $z=0$, in the 0.2-200 GeV energy range. Top panel: $\gamma$-emission for clusters in the CUR1 box, assuming CR-spectra of  $s=2.0$ or $s=3.0$. Centre: $\gamma$-ray emission from clusters in the CUR2 runs, for different models of gas physics. Bottom: $\gamma$-ray emission from clusters in our non-radiative CUR2 run, for runs with different acceleration efficiency of CRs. The gray symbols are the upper limits from the FERMI catalog in the same energy range.}
  \label{fig:gamma}
\end{figure}

\subsection{Hadronic $\gamma$-ray emission from simulated cluster samples}
\label{subsec:result_cosmo}

The hadronic emission from the CRs population of each simulated cluster is computed following the standard formalism of \citet[][and]{pe04,donn10}, with the only difference that for the hadronic cross-section we use the parametrisation of the proton-proton cross section given by \citet{2006PhRvD..74c4018K}, as in  \citet{2013A&A...560A..64H}. A recent review of the method is given in \citet{va15relics}. Our 2-fluid formalism cannot follow particle spectra and therefore we have to guess a fixed spectral energy distribution of CRs in the simulated volumes. We consider here the large $0.2-200$ GeV energy range to be less sensitive to the exact spectral energy distribution of CRs, which is not directly simulated in our method.
The first panel of Figure \ref{fig:gamma} shows the distribution of the predicted emission as a function of the cluster mass for the  non-radiative clusters of the CUR1 run, where we computed the hadronic emission from each cluster for the cases of a fixed 
$s=2.0$ (corresponding to a $\gamma$-ray spectrum of $s_\gamma \approx 2.5$ at high energy{\footnote{$s_{\gamma}$ is the spectrum of the $\gamma$-ray emission and is given by $s_{\gamma}=4(s-1/2)/3$ \citep[e.g.][]{pe04}.}})  and $s=3.0$ ($s_\gamma \approx 3.33$). 
We notice that here we assume slightly steeper spectra compared to \citet[][]{2010MNRAS.409..449P}, who found nearly universal energy spectra compatible with $s \approx 2.3$ for most of the CR-energy range. In our case, these estimates follows from the distribution
of CR-energy injected by merger shocks seen in these simulations \citep[][]{va09shocks,va10kp,va11comparison}, and are also confirmed by our tracer-based modelling of 
the following Section (Sec.~\ref{subsec:result_amr}).
In all cases, considering the large energy range used here to compare with FERMI data, the effect of the spectral shape is not big in our predictions and the total photon flux is only varied by a $\sim 50\%$ going from $s=2.0$ to $s=3.0$, not a big effect. 
Our predictions are compared to the observed upper limits from the FERMI satellite, within the same energy range, obtaining by converting each limits on the received photon flux into a limit on the absolute luminosity at the distance of each object. For $\sim 50 \%$ of simulated clusters the predicted emission is at the level or above the upper limits from FERMI observations.\\
The effect of different prescriptions for gas physics or CR physics is shown in the lower panels of Figure \ref{fig:gamma}. Radiative feedback worsens the comparison with
observations, by producing typically denser gas cores (CUR2 run with low redshift feedback) or by increasing the number of shocks connected to AGN feedback (CUR2 run with
high redshift feedback). 
The use of the (higher) acceleration efficiency assumed in \citet{kj07} obviously produces an even higher hadronic flux. 
The approximated version of the  \cite{2014ApJ...783...91C} acceleration model significantly reduces the number of objects above FERMI limits, but does not to entirely solve the problem as still $\sim 10-20 \%$ of simulated objects are above FERMI limits.
We found that only by limiting the overall acceleration efficiency of shocks to $\eta=10^{-3}$ for all Mach numbers, the predicted hadronic emission goes below the FERMI limits in the non-radiative case, while the fraction of clusters that is inconsistent with FERMI limits is now limited to $\leq 10 \%$ in the cooling+AGN feedback case. In the latter case, these high $\gamma$-ray emitters are the 
densest CC-like objects in the volume. \\

\begin{figure*}
    \includegraphics[width=0.33\textwidth]{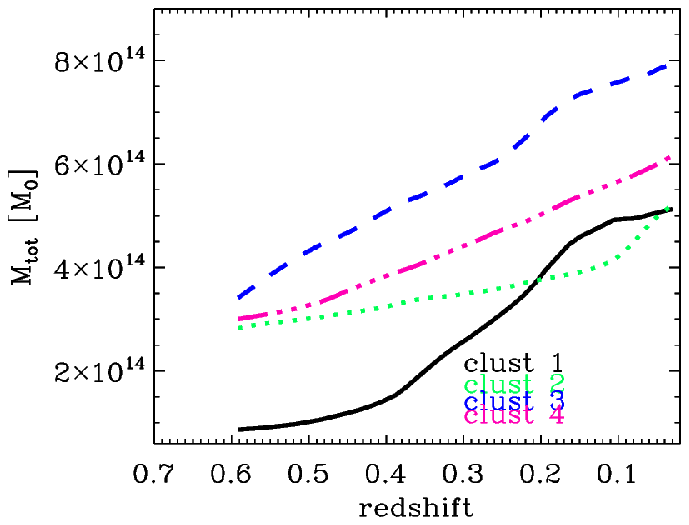}
      \includegraphics[width=0.33\textwidth]{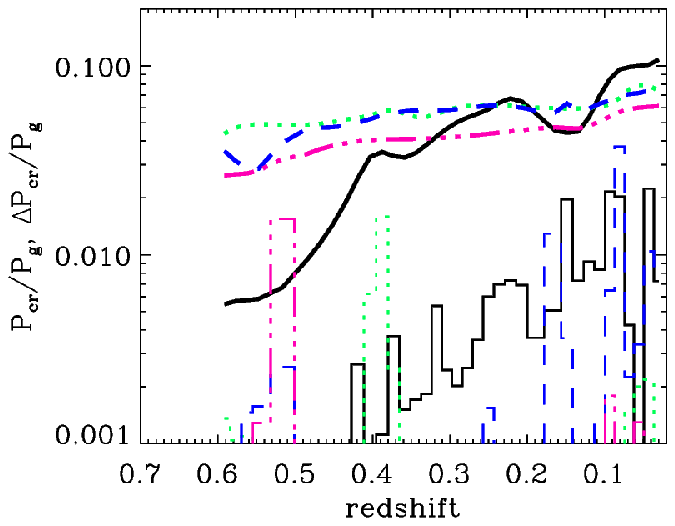} 
    \includegraphics[width=0.33\textwidth]{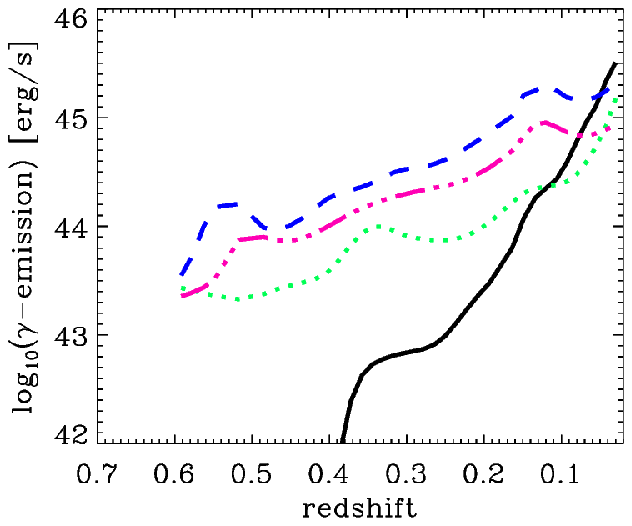}
    \caption{Evolution of four clusters with a final mass $\geq 5 \cdot 10^{15} M_{\odot}$ in the CUR1 volume. From left to right the image shows: a) the enclosed total mass with fixed $6^3 \rm Mpc^3$ comoving volumes; b) the total CR to gas pressure ratio (thick lines) and the relative CR pressure increment snapshot by snapshot (think lines); c) the total hadronic $\gamma$-ray emission from the same volumes.}
  \label{fig:4cluster}
\end{figure*}

All models vary with mass in that the hadronic emission scales as $\epsilon_{\gamma} \propto M^{5/3}$, with normalisation varying with the physical model and a $\sim 1-2$ dex scatter around the mean relation that increases going to the lowest masses. While the $\propto M^{5/3}$ relation is in line with earlier results by \citet{2008MNRAS.385.1242P} and \citet{2010MNRAS.409..449P}{\footnote{The $\sim 5/3$ exponent can be understood if the average CR to gas energy ratio is a constant, in which case the total $\gamma$-emission scales as the thermal gas energy of the cluster, $\epsilon_{\gamma} \propto E_{\rm g} \int_V n X$, and $E_{\rm g} \propto M^{5/3}$ in virialised clusters \citep[e.g.][]{va06}.}}, the level of scatter we observe is significantly larger.\\ 
To understand the origin of this large scatter we focus on 4 clusters in the largest CUR1 box, with large final masses ($\geq 5 \cdot 10^{14} M_{\odot}$) but different dynamical histories. 
For each of these objects we analyzed 50 snapshots equally spaced in time from $z=0.6$ to $z=0$, and extracted the mean properties of gas and CRs within fixed comoving volumes centred on each object. Figure \ref{fig:4cluster} gives the evolution of the total enclosed mass, of the CR to gas pressure ratio (including the CR pressure ratio generated by the injection of CRs over each single snapshot), and of the enclosed $\gamma$-ray emission.
The evolutionary tracks of these clusters show that the total cluster mass is not the only variable that sets the $\gamma$-ray emission, but also the mass accretion rate induces significant scatter on top of the global $\epsilon_{\rm \gamma} \propto M^{5/3}$ relation. For example, cluster 1 has the final smallest mass in this sample, but the highest hadronic emission at $z=0$, because it has been rapidly assembled during $z=0.4-0.2$ in a merger event. This led to an increase of the number of shocks, to enhanced injection of CRs and to a nearly $\sim 10^4$ times increased $\gamma$-ray emission from $z=0.4$ to $z=0$. On the other hand, the hadronic emission of more
massive but more relaxed clusters 3 and 4 is only increased by $\sim 10^2$ from $z=0.6$ to $z=0$. Cluster 2 experiences a major merger late in its evolution ($z \leq 0.1$), which leads to a $\sim 10$ times increase of hadronic emission over the last  $\sim 1.5$ Gyr.
These different evolutionary paths show that the hadronic emission in our clusters can vary by a few orders of magnitudes during the typical crossing time of clusters, due to the enhanced injection of CRs and due to the gas compression of the ICM during mergers.  Cooling and AGN feedback further add a source of scatter to the relation with the host cluster mass.
We further elaborate on the variance of our results compared to \citet{2008MNRAS.385.1242P} and \citet{2010MNRAS.409..449P} in Sec.\ref{sec:discussion}.

\subsection{Hadronic $\gamma$-ray emission from MACSJ1752}
\label{subsec:result_amr}

\begin{figure*}
    \includegraphics[width=0.495\textwidth]{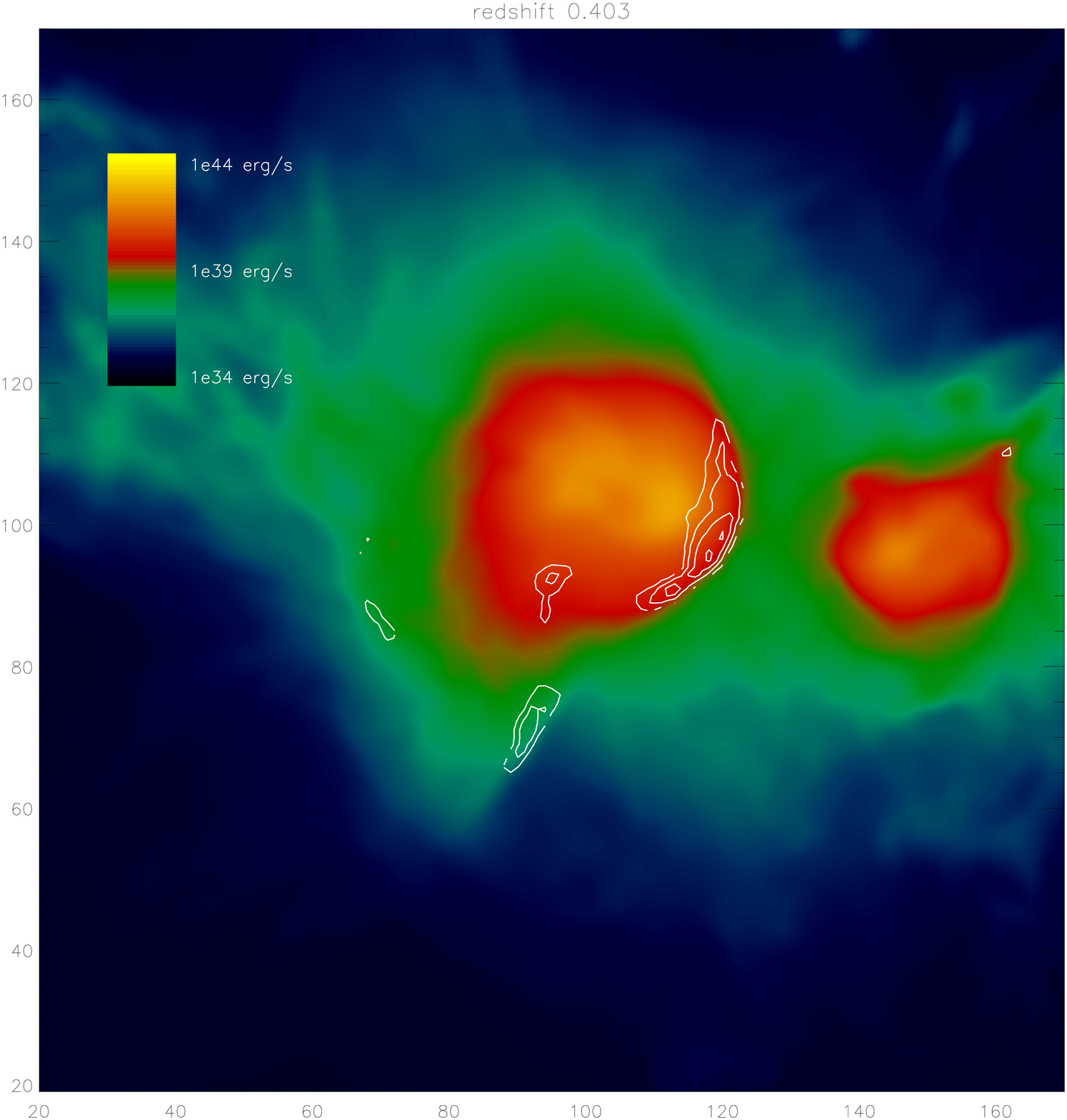}
      \includegraphics[width=0.495\textwidth]{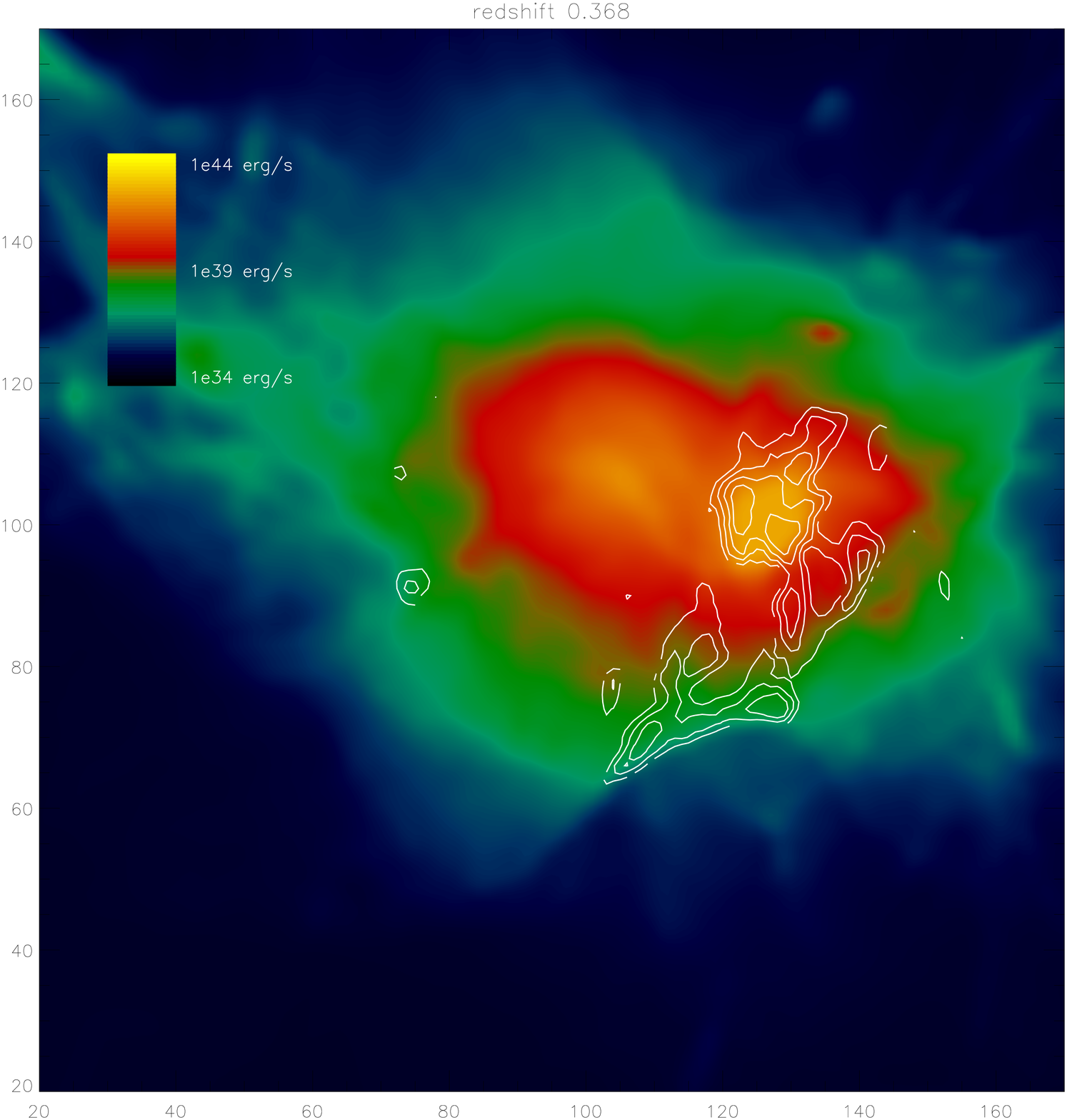}  
      \includegraphics[width=0.495\textwidth]{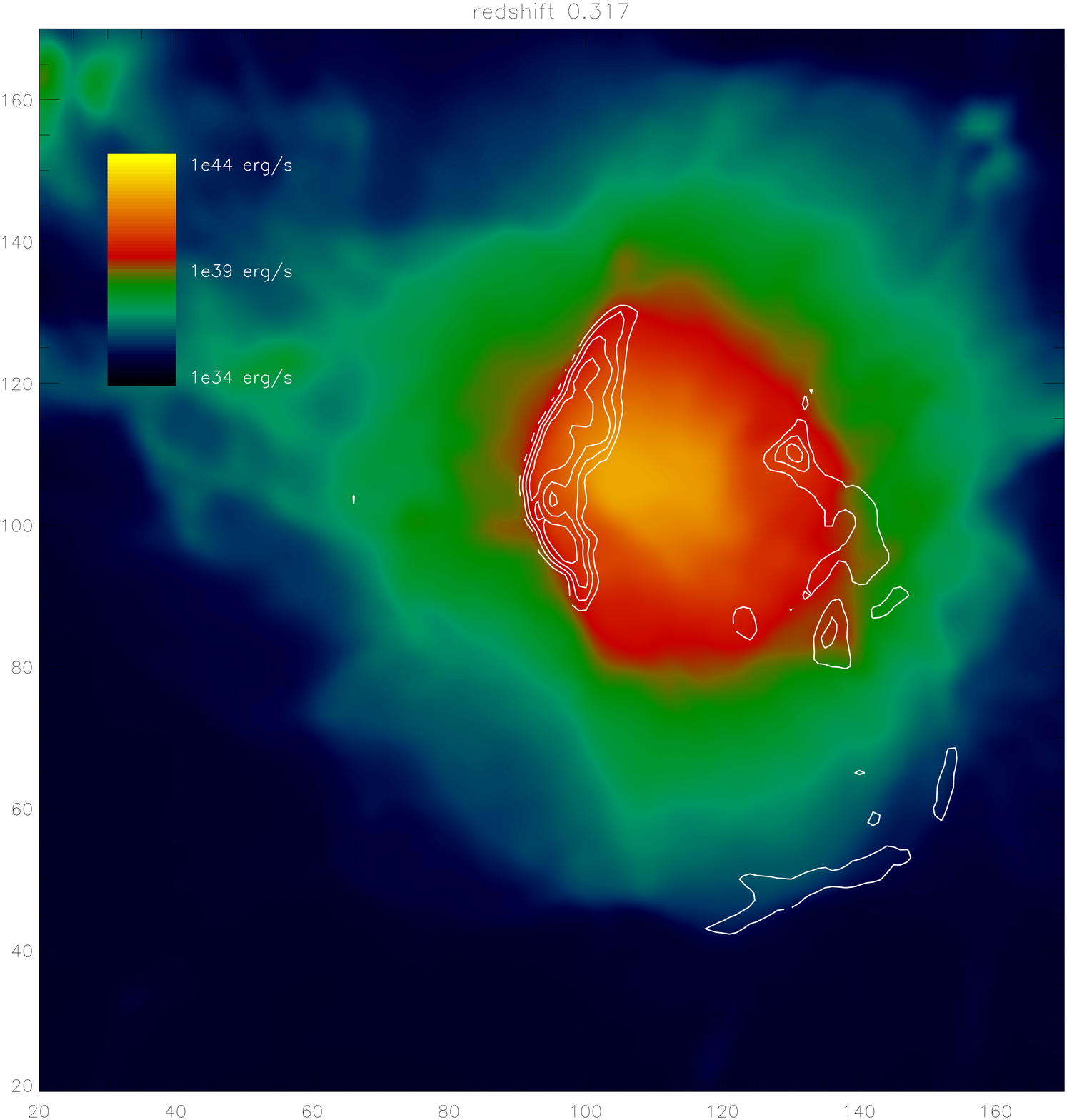}  
         \includegraphics[width=0.495\textwidth]{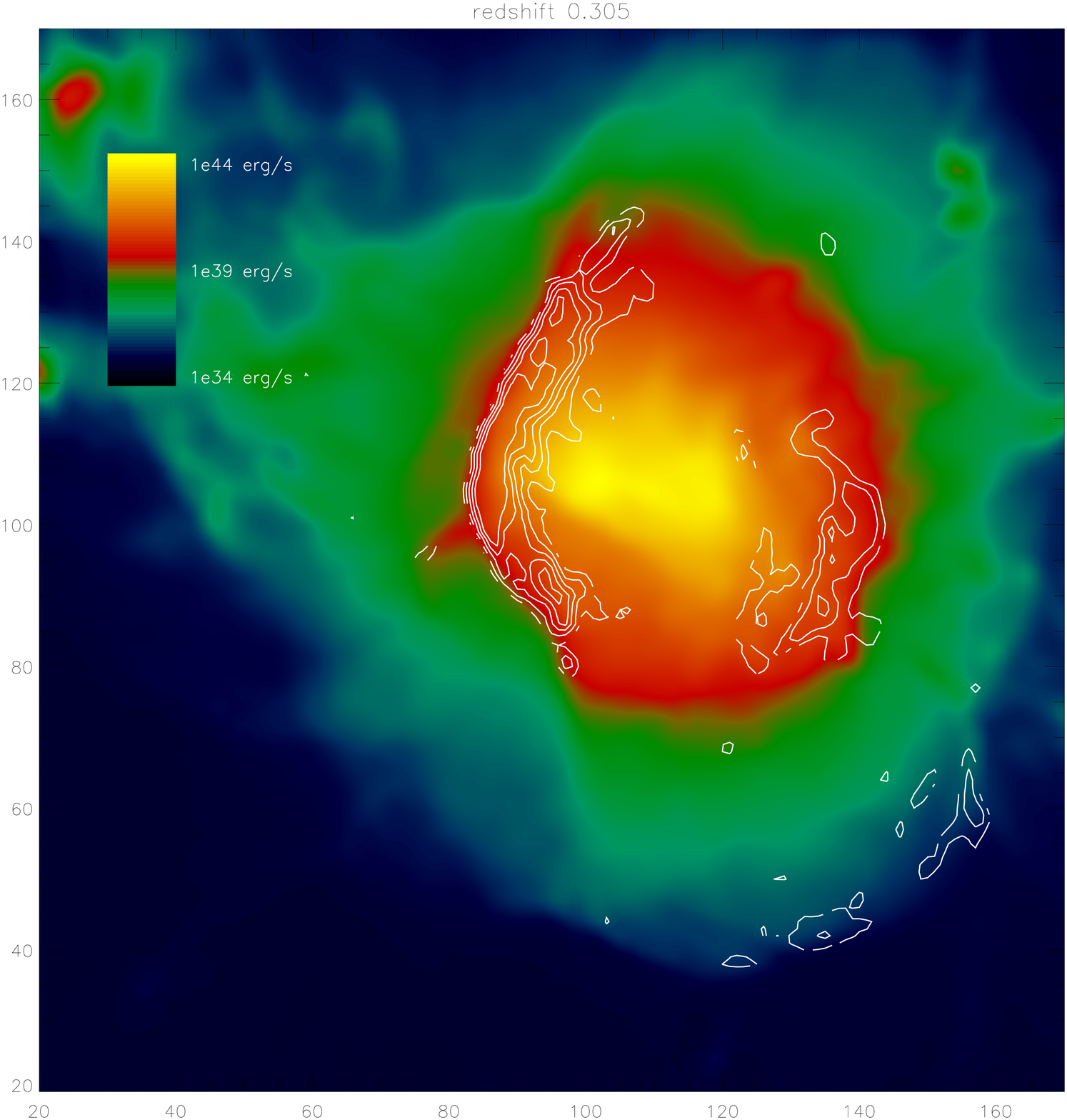}  
                         \caption{Simulated merger sequence for a $\sim 10^{15} M_{\odot}$ cluster using AMR, with X-ray emission in colours and radio emission in white contours. The approximate epoch of the observed merger in MACSJ1752.0+0440 is $z \approx 0.3$. Each image is $5 \times 5 ~\rm Mpc^2$ (comoving) across. The contours are equally spaced with $\sqrt 2$ multiples of the radio emission, starting from $\approx 10^{23} \rm erg/(s ~Hz)$ per pixel.}
  \label{fig:merger}
\end{figure*}

We finally focus on the resimulations of a major merger event leading to X-ray and radio morphologies similar to the observed
cluster MACSJ1752.0+0440, which we already studied in  \citet[][]{2012MNRAS.426...40B}.
Figure \ref{fig:merger} gives the merger sequence ($0.4 \geq z \geq 0.25$) for the simulated cluster,
showing the projected X-ray brightness  and the radio emission using the formalism by
\citet{hb07}. Since the magnetic field is not included in our simulation, we assume that the energy in the magnetic field is everywhere $1/\beta=1/100$ of the thermal gas energy within the cell, which gives $\sim \rm \mu G$ fields in this case. For the electron acceleration efficiency at shocks, we assumed that this is $10^{-3} \times \eta(\mathcal{M})$, where $\eta(\mathcal{M})$ is the acceleration efficiency by \citet{kr13}.\\

The total virial mass of the system after the collision  is $\approx 0.65 \cdot 10^{15} \rm M_{\odot}$, while the mass ratio of the merger is $M_1/M_2 \sim 1.6$. At $z \sim 0.3$ the shock propagating downstream
of the main cluster progenitor has $\mathcal{M} \approx 4.5$ and produces a total radio emission of $P_{\rm radio} \sim 10^{26} \rm erg/(s ~ Hz)$ at 325 MHz over an extent of $\sim 1.5 ~\rm Mpc$, similar to observations.  Downstream of the smaller cluster progenitor, also a second radio relic  $\sim 700 ~\rm kpc$ wide and with a total power $P_{\rm radio}  \sim 3 \cdot 10^{25} \rm erg/(s ~Hz)$ is formed. 
The double relic configuration observed in MACSJ1752.0+0440 at $z \approx 0.3$ should be produced  $\sim 0.8$ Gyr after the central collision of the cores, and we use this epoch also to constrain the 
acceleration scenario of CRs based on the lack of $\gamma$-ray emission. \\

\begin{figure}
    \includegraphics[width=0.49\textwidth]{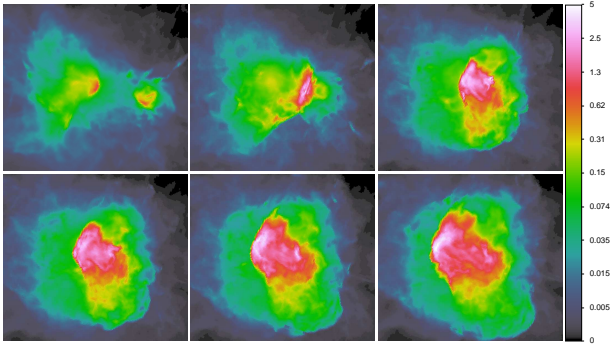}
 \caption{Merger sequence for our simulated versions of the cluster  MACSJ1752.0+0440 using AMR for the same area of Fig.\ref{fig:merger} and for the epochs of $z=0.403$, $0.368$, $0.317$, $0.305$, $0.294$ and $0.281$. The colours show the total projected CR-energy in code units.}
 \label{fig:merger2}
\end{figure}

The evolution of the projected CR-proton energy simulated with our two-fluid method in {\enzo} is given in Fig.\ref{fig:merger2}, for the \citet{kr13} injection model. The late energy budget of CRs in the innermost cluster regions is clearly dominated by the injection through the $\mathcal{M} \sim 4$ shock launched by the major merger at $z \approx 0.3$.
However, on larger scales the CR-energy mostly results from the cumulative activity of previous formation shocks and can be considered as a background CR-energy distribution present regardless of the last major merger.  We notice that, compared to our first study in  \citet[][]{2012MNRAS.426...40B}, the effect of the additional  CR-pressure in this set of simulations results into a 
less satisfactory match of the simulated X-ray and radio emission compared to the observation. However, several significant features as the elongated a disrupted inner X-ray structure, the 
emergence of the double relic and the off-set of the least powerful relic respect to the merger axis are well reproduced. 
Given that the masses and the relative distances between the gas cores and the relics allow a reasonable reconstruction of the merger scenario, we consider
this system an interesting case to test CR acceleration models against the FERMI limits for this system (Sec.\ref{subsec:fermi_macs}).\\

 While our two-fluid formalism only allows us to monitor the total
CR-energy in each
cell, with the complementary use of passive tracer particles we can better
follow the spatial evolution of CRs in the ICM \citep[e.g.][]{va11entropy}.
To this end, for one AMR run we saved $\sim 50$ equally time-spaced
snapshots from $z=1$ to $z=0$ and we evolved the trajectory of passive tracers using the
output 3-dimensional
velocity field. The underlying assumption is that CRs can be treated as
frozen into
the gas as their spatial diffusion in negligible on these scales ($\sim 34
~\rm
kpc$).  We injected in total $\approx 6 \cdot 10^6$ tracers at $z=1$, with
initial
distribution proportional to the gas density in the {\enzo} run.
Their positions were updated at every time step with explicit time integration and using the same Courant condition of the {\enzo} simulation. The tracer velocity was computed from the grid values at the tracer position using a
CIC interpolation that includes a turbulent correction term to better model the mass flux.  For each
tracer we recorded the CR energy read from the grid, and evolved the spectral
index of  the CR-energy of each tracer integrating over all injected populations of
CRs, assuming an injection spectrum of 
$s=2(\mathcal{M}^2+1)/(\mathcal{M}^2-1)$.  In
the case of multiple shocks hitting a tracer, the final CR-spectrum kept the
smallest slope given by the time sequence of shocks \citep[e.g.][]{kr10}, while the
normalisation was updated based on the CR-energy, computed from the
grid. With this approach we can monitor the spatial evolution of CR spectra in the
cluster and have a better modelling of the resulting $\gamma$-ray
emission.  More
details on the tracers algorithm will be
given in a  forthcoming paper, by Wittor et al. (in preparation).

\begin{figure}
    \includegraphics[width=0.45\textwidth]{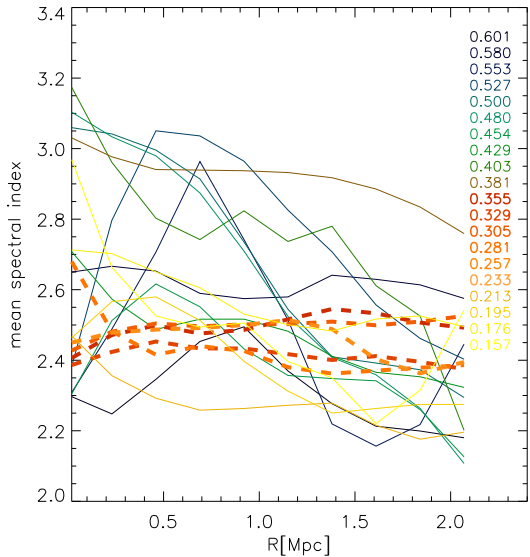}
    \caption{$\gamma$-ray emission weighted radial distribution of the spectral index of CR-energy in the AMR resimulation of  MACSJ1752 for different redshifts (marked in different colours). The thick dashed lines marked the epochs closer to the observed radio emission.}
  \label{fig:prof_CR_macs}
\end{figure}

Figure \ref{fig:prof_CR_macs} shows the average $\gamma$-ray emission weighted profile of CR-spectra for this cluster at different redshifts, with the thick dashed lines marking the epochs around the merger event.
The average spectrum of CRs is initially steep in the cluster centre, $s \approx 3$ and flatter in the outskirts, $s\approx 2.2 $, mirroring the
typical distribution of shocks in clusters, which are weaker in the centre and stronger outwards \citep[e.g.][]{va11comparison}. 
Later on, with the progress of the merger the spectrum flattens everywhere as the $\mathcal{M} \sim 4-5$ shock sweeps the intracluster volume. 
At the approximate epoch of the X/radio observation, the average spectrum has a flat distribution around $s \sim 2.4-2.5$, with significant fluctuation patches spread over the cluster volume.  

\begin{figure}
    \includegraphics[width=0.45\textwidth]{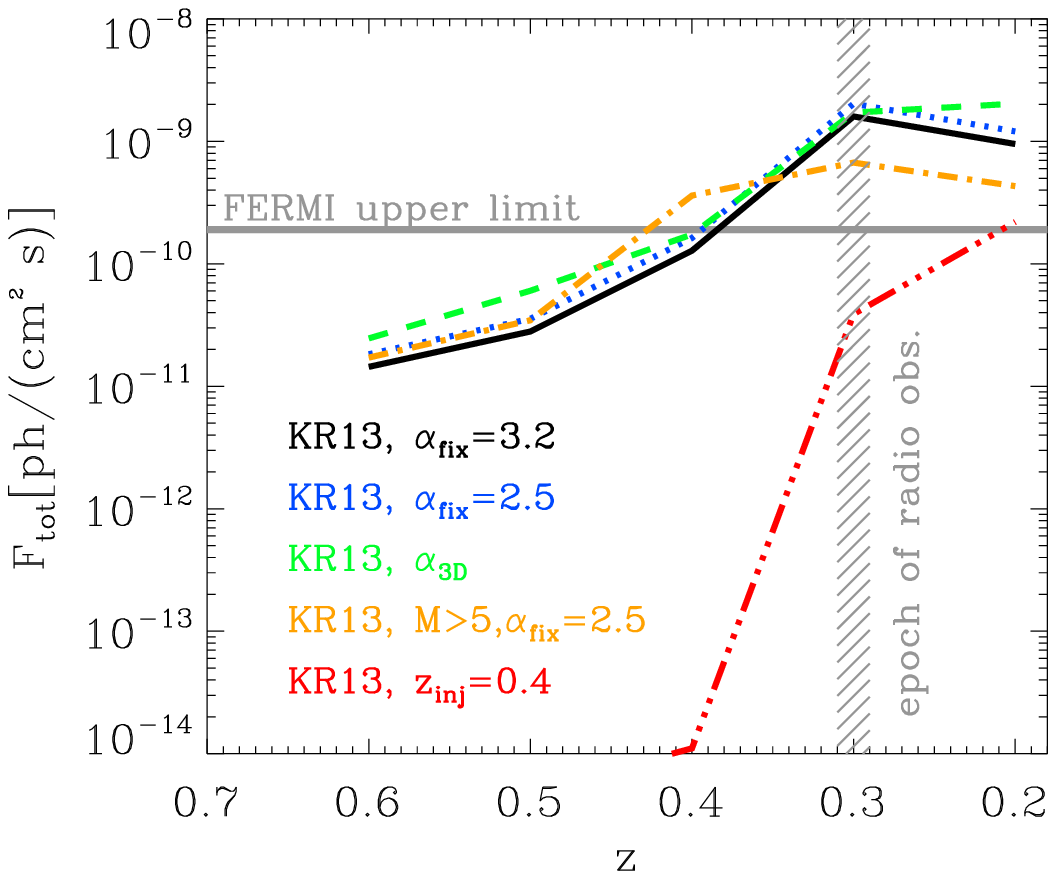}
    \caption{Evolution of the total hadronic $\gamma$-ray emission in the $0.2-100$ GeV energy range for the various resimulations
    of cluster MACSJ1752. The vertical hatched region shows the epoch of the observed X-ray/radio configuration, while the horizontal line marks the upper limits from FERMI on the
    hadronic emission from this cluster.}
  \label{fig:second}
\end{figure}

As a final result, the evolution of  the predicted $\gamma$-ray emission for a few scenarios for CRs is given in Fig. \ref{fig:second}. 
Here we show the hadronic emission obtained using fixed CR-spectra of $s=2.5$ or $s=3.2$, or directly using the 3D spectral information
from the tracers. Despite fluctuations of order $\sim 2-3$ in the total emission, the FERMI limits
derived in Sec.~\ref{subsec:fermi_macs} are in all cases exceeded by a factor $\sim 10$  at the epoch of the radio observation of MACSJ1752.0+0440. 
It is not only the major merger that is responsible for the boost of the $\gamma$-emission: in an additional run assuming the \citet{kr13} model we only allowed the injection/re-acceleration of CRs by $\mathcal{M}\geq 5$ shocks. In this case the hadronic emission is greatly reduced compared to all previous models, yet the FERMI limits are still exceeded by a factor $\sim 2-3$ at $z=0.3$.
Finally, as a simple test we rerun the AMR simulation by imposing that the acceleration of particles only begins at $z=0.4$, i.e. when the major merger begins.
The hadronic emission ramps up during the merger, yet the predicted emission at $z=0.3$ is below the constraints by FERMI. This confirms that while the injection of CRs during a merger can significantly boost the hadronic emission from the ICM, the problem of simulated models to full fill FERMI constraints more generally stems from the assumed CR-enrichment across the full structure formation process. 
Our findings here for MACSJ1752.0+0440 are also consistent with the semi-analytical model we presented in \citet{va15relics}, where the predicted hadronic emission resulting from the major merger is $\sim 10 \%$ of the FERMI limit for this object. However, in the case of a few closer systems with relics such as A3367, ZwCLJ2341, A754 and A2256, the limits by FERMI are challenged even by considering the last merger only \citep[see][for more details]{va14relics,va15relics}.

\section{Discussion}
\label{sec:discussion}

Both our non-radiative and radiative runs produce galaxy clusters with a budget of CRs in tension with the latest FERMI limits \citep[e.g.][]{fermi13}, suggesting the necessity of a revision of the acceleration efficiency from DSA in cosmic shocks. 
These conclusions are in line with the results of our previous semi-analytical modelling of double relics \citep[][]{va14relics,va15relics} and put typically stronger constraints on the acceleration of CRs by cosmic shocks compared to previous works based on simulations \citep[e.g.][]{ry03,ka07,pf07,2010MNRAS.409..449P,scienzo,va13feedback}. 

We can briefly discuss possible explanations for this finding. First, the real efficiency from diffusive shock (re)acceleration of protons by weak shocks ($\mathcal{M} \leq 5$) is  $\sim 10-100$ times below the current estimates from DSA. This is reasonable as the latest results of particle-in-cell and hybrid simulation of cosmic shocks are revising downwards the estimates of CR-protons acceleration efficiency \citep[][]{ca14a,guo14a}. Our test based on the \citet{ca14a} acceleration model shows that the tension with FERMI is alleviated if the overall DSA acceleration efficiency is revised downwards assuming that only quasi-parallel shocks efficiently accelerate CR-protons, and that the acceleration efficiency is $\sim 1/2$ of what derived in \citet{kr13}.
However, in this case the acceleration mechanism in radio relics must be different from DSA, due to the too large electron-to-proton ratio ($K_{\rm e/p}>10^{-2}$) needed to explain the data \citep[][]{va14relics,bj14,va15relics}. 
The shock-drift-acceleration mechanism may indeed lead to this result, as shown by \citet{guo14a,guo14b}. However, due to computational limitations it was impossible to reach the full FERMI I acceleration regime, and therefore the acceleration efficiency of electrons is not yet well constrained in this scenario.

On the other hand, aged relativistic electrons that are reaccelerated by shocks can solve the problem \citep[e.g.][]{pinzke13,2015ApJ...809..186K}, but only if the electrons have been injected by leptonic-dominated jets from radio galaxies (or other leptonic scenarios). If instead the aged electrons are assumed to be the result of previous injection by cosmic shocks, the problem is even exacerbated \citep[][]{va14relics,va15relics}.
In addition, the acceleration efficiency from DSA might be a steep function of the up-stream magnetisation level, i.e. for typical 
intracluster or intergalactic magnetic fields below some given threshold the acceleration efficiency might drop. This would limit the 
injection of CRs to some fraction of the intracluster volume, where the magnetic field is amplified, and to the fraction of the cosmic time
where cosmic magnetic fields have grown enough \citep[e.g.][]{br05,do08,wi11,va14mhd}. However, assessing the limiting magnetic field needed to produce DSA in these regimes is difficult, also because upstream magnetic fields might be always amplified by a CR-driven dynamo \citep[][]{2012MNRAS.427.2308D,2013MNRAS.tmp.2295B,2015PhRvL.114h5003P}. 
While in this work we could only explore the impact of magnetic fields in a very crude way (CS14 model) by assuming they are randomly oriented in space, we defer to future work with MHD simulations (Wittor et al., in prep), that will exactly address this issue.

Finally, we discuss the main limitations of our modelling and the comparison with the literature.
\begin{itemize}
\item {\it Comparison with SPH results.}  Our distribution of the CR-to-gas pressure ratio is at variance with SPH \citep[][]{2010MNRAS.409..449P}, who found a much tighter relation between the host cluster mass.  We find instead that mergers boost the ratio up to $\sim 1-2$ orders of magnitude. This is mostly due to the steeper acceleration efficiency model we assume here \citep[][]{kr13}. For example, our acceleration efficiency goes from $\approx 10^{-5}$ at $\mathcal{M}=2$ to $\approx 10^{-2}$ at $\mathcal{M}=3$. On the other hand, in the model by \citet{2007A&A...473...41E} the acceleration efficiency goes from $\approx 0.01$ to $\approx 0.2$ in the same range of Mach numbers, i.e. their acceleration efficiency is $\sim 10-10^2$ times larger than ours. Therefore, our model for CRs is more sensitive to the rare $\mathcal{M} \geq 3$ shocks driven by mergers, while in \citet[][]{2010MNRAS.409..449P} the injection of CRs is overall more constant over time because also $\mathcal{M} < 3$ shocks inject significant CRs. 
Additionally, the way in which gas (and the frozen-in CRs) is transported in the central core of clusters simulated with grid and SPH methods 
is known to be different \citep[e.g.][]{1999ApJ...525..554F,2010MNRAS.401..791S,va11entropy,va11comparison,2012MNRAS.424.2999S,2015MNRAS.454.2277S} and this can further amplify differences in the budget of CRs in the cluster centre. 

\item {\it Comparison with grid results}. Compared to the earlier work by \citet{mi01} and \citet{2003MNRAS.342.1009M} we improve on the spatial resolution by $\sim 2-4$ times and the total number of simulated clusters by $\sim 10^2$ times. The hydro scheme we applied is also different, and the CRs have a dynamical impact on the gas evolution, while they are passive in  \citet{mi01} and \citet{2003MNRAS.342.1009M}. Our main findings however, including the cluster to cluster scatter, are basically in line with these previous results.

\item{\it Comparison with semi-analytical results.} The results of this work are overall in agreement with our recent modelling of CRs acceleration in double relics using semi-analytical methods \citep[][]{va14relics,va15relics}. There we limited our analysis to the (re)acceleration of electrons and protons by the merger shocks which should be responsible for the observed radio emission, and therefore we had to neglect the previous enrichment of CRs during structure formation. Even in this case, we find that overall the acceleration efficiency of protons must be  limited to $\leq 10^{-3}$ for the range of Mach numbers inferred by radio spectra, $\mathcal{M} \sim 2-5$. 

\item {\it Spatial resolution.}  Our spatial resolution is the result of a compromise between the necessary detail on shocks and on the need of sampling large volumes. Earlier works \citep[e.g.][]{ry03,sk08,va09shocks,va11comparison} show that a spatial resolution of $\sim 200-300 ~\rm kpc$ is enough to properly resolve the thermalisation by cosmic shocks. Our analysis of these simulated volumes  \citep[][]{scienzo14} also shows that at the resolution of $\sim 100-200~\rm kpc$ the global properties of CRs are converged.  Additional resolution tests are given in the Appendix. 

\item {\it Cooling and AGN feedback.}  While we do not find a significant {\it direct} impact of AGN feedback and CR enrichment for the large scales of interest here, the cumulative effect of the cooling-feedback interplay can change the  thermodynamical structure of the ICM and affect the CR-to-gas ratio and the $\gamma$-ray emission. Modelling AGNs in cosmological simulations is still challenging \citep[][]{2006MNRAS.366..397S,2009MNRAS.400..100S,dubois10,mcc2010}, and so is the complex interplay between AGN injected CRs and the surrounding gas \citep[][]{alek12,2013ApJ...779...10P,va13feedback}. However, our simulations show that including cooling and feedback {\it worsens} the comparison with FERMI limits. Therefore our main findings on the too large acceleration of CRs should be conservative against more complex models for the gas physics. 

\item {\it Limitations of CR physics}. Our runs do not include all possible interactions between CRs and gas and we assume that CRs do not diffuse neither stream out of the gas distribution. However, in \citet{va13feedback} we tested that including CR-losses does not affect the distribution of CRs outside of cluster cores, while most of the $\gamma$-emission is mostly produced on larger scales. Additional mechanisms of (re)acceleration of CRs, such as turbulent reacceleration \citep[e.g.][]{bj14}, reconnection \citep[e.g.][]{2015ASSL..407..311L} and injection by AGN and supernovae \citep[e.g.][]{1996SSRv...75..279V} can only increase the budget of CRs estimated in our simulations.
The spatial diffusion of CRs can modify the distribution simulated by our 2-fluid method  \citep[][]{bbp97}. Yet again, this cannot significantly affect our predicted $\gamma$-ray emission, which is spread across $\sim \rm Mpc$ scales, while the diffusion of $\sim 1-10 ~\rm GeV$ protons over Gyrs can only affect $\sim 10-10^2 ~\rm kpc$ scales. Finally, 
the detailed calculation by \citet{2013MNRAS.434.2209W} showed that even in a scenario in which CRs can stream out of the gas distribution faster than the Alfv\'{e}n speed, the $\gamma$-ray emission is not affected in the energy range probed by FERMI, but only at much higher energies ($E=300-1000$ GeV).
\end{itemize}

\section{Conclusions}
\label{sec:conclusions}

We simulated the acceleration of CRs by structure formation shocks, and focused on the observable $\gamma$-ray outcome of CRs in galaxy
clusters. Our suite of simulations has been designed to simulate several realistic scenarios for CRs and gas physics, and to enable the testing for large complete samples of galaxy clusters \citep[][]{scienzo14}. 

\begin{itemize}

\item All tested CRs (re)acceleration models based on DSA predict a level of hadronic emission inconsistent with $\gamma$-ray observations.  A significant fraction of our simulated clusters ($\sim 25-50 \%$ depending on the model) produce  $\gamma$-ray emission above the upper limits reported by FERMI \citep[][]{ack10,2013A&A...560A..64H,fermi13,fermi14,2014ApJ...795L..21G,zand14}.This result is robust against the variations in the gas physics that we have tested.
   
 \item Regardless of the models (with the exception of the unrealistic gas model producing overcooling within halos) we find that the average radial profile of CRs to gas  pressure, $X(R)$, is well described by a simple 2nd order polynomial expression (Eq.1). In all models the profiles are very flat 
  for $R \geq 0.2 R_{\rm vir}$, and have a central normalisation $X_0$ depending on the acceleration model. This average profile can be useful for the inversion of FERMI data of clusters \citep[e.g.][]{fermi14}.

\item Clusters of similar mass have a total pressure ratio $X$ that varies up to a factor $\sim 10-10^2$ depending on their dynamical state. In non-radiative simulations, the main source of scatter is caused by mergers, which inject new CRs and boost the hadronic emission. In radiative simulations, cooling and feedback can also increase the CR-to-gas pressure ratio by lowering the gas temperature and by triggering stronger shocks from the innermost cluster regions of CC-like clusters. 

\item The different average CR-to-gas pressure ratio in clusters with similar mass, due to their different dynamical history, suggests that the present FERMI limits are actually constraining the total budget of CRs in the average cluster population to a deeper level than usually assumed. Indeed, the non-detection of hadronic emission mostly puts a constrain on the budget of CRs in the active systems, which must be limited to $X \sim 1\%$ to be consistent with FERMI limits \citep[][]{2013A&A...560A..64H,fermi13,fermi14,2014ApJ...795L..21G}. However, given the range of scatter in $X$ across our simulated clusters, we conclude that in more relaxed systems this ratio must be even lower, $X \sim 0.1\%$, i.e. nearly ten times below the average limit given by the modelling of FERMI stacking \citep{2013A&A...560A..64H,fermi13}.

 \item This result  suggests that the (re)acceleration efficiency assumed for CRs must be significantly scaled down compared to all models tested here, which are derived from diffusion-convection simulations of DSA \citep[][]{kj07,kr13}. Revising the acceleration efficiency by \citet{kr13} downwards based on the latest results of hybrid simulations by \citet{ca14a} is found to limit the tension with FERMI data to a few objects.  A {\it fixed} proton (re)acceleration efficiency $\eta=10^{-3}$ at all shocks produces hadronic emission from clusters below the present upper limits from FERMI. In this case the central CR to gas pressure ratio is $\approx 0.6\%$ and reaches $1\%$ within the virial radius. The energetics of CRs in clusters is largely dominated by $\mathcal{M} \sim 2-5$ shocks, this turns into a requirement for the acceleration efficiency in this Mach number range. However, also the acceleration efficiency by stronger external shocks (usually assumed in the range $\eta \sim 10 \%$) must be revised, as our test in Sec.~\ref{subsec:result_amr} shows. This result is in agreement with the independent modelling based on semi-analytical mergers discussed in \citet{va15relics}, where also a $\sim 10^{-3}$ acceleration efficiency of protons at shocks was found to be necessary to reconcile with the lack of detected hadronic emission from clusters hosting radio relics. 

\item These results stregthen the problems we pointed out in a related series of semi-analytical paper focusing on double relics \citep[][]{va14relics,va15relics}, where we reported that in order to reconcile the shock-acceleration model of relics with the absence of $\gamma$-ray emission from protons, a electron-to-proton injection ratio ($\gg 10^{-2}$) larger than canonic DSA must be assumed. A promising solution
to this problem might come from the latest work on particle acceleration given by particle-in-cell simulations \citep[][]{guo14a,guo14b}, which found efficient pre-acceleration of relativistic electrons and very little proton injection by $\mathcal{M} \leq 5$ shocks, due to shock-drift-acceleration (SDA). In addition, hybrid \citep[][]{ca14a,ca14b} or PIC \citep[][]{2015PhRvL.114h5003P} simulations of proton acceleration by stronger shocks reported large variations of the injected CR-energy depending on the shock obliquity with the upstream magnetic field. 

\end{itemize}
In conclusion, this study shows that cosmological simulations joined with $\gamma$-ray observations can be used as a laboratory to study the acceleration of cosmic rays by weak cosmic shocks, in a regime complementary to supernova remnants and solar wind shocks.

\section*{acknowledgments}
This work was strongly supported by computing resources from the Swiss National Supercomputing Centre (CSCS) under projects ID ch2 and s585 in 2014-2015. FV acknowledges personal support from the grant VA 876/3-1 from the Deutsche Forschungsgemeinschaft. FV and MB also acknowledge partial support from the grant FOR1254 from the Deutsche Forschungsgemeinschaft. We acknowledge the use of computing resources under allocations no. 7006 and  9016 (FV) and 9059 (MB) on supercomputers at the NIC of the Forschungszentrum J\"{u}lich.  We acknowledge the use of the online "Cosmology Calculator" by \citet{2006PASP..118.1711W}. We thank K. Dolag for very fruitful discussions at an early stage of this work, and our referee H. Kang for very useful comments on the manuscript.

\bibliographystyle{mnras}
\bibliography{franco}

\appendix

\section{Effects of resolution and additional physics}
\label{appendix1}
\begin{figure}
     \includegraphics[width=0.24\textwidth]{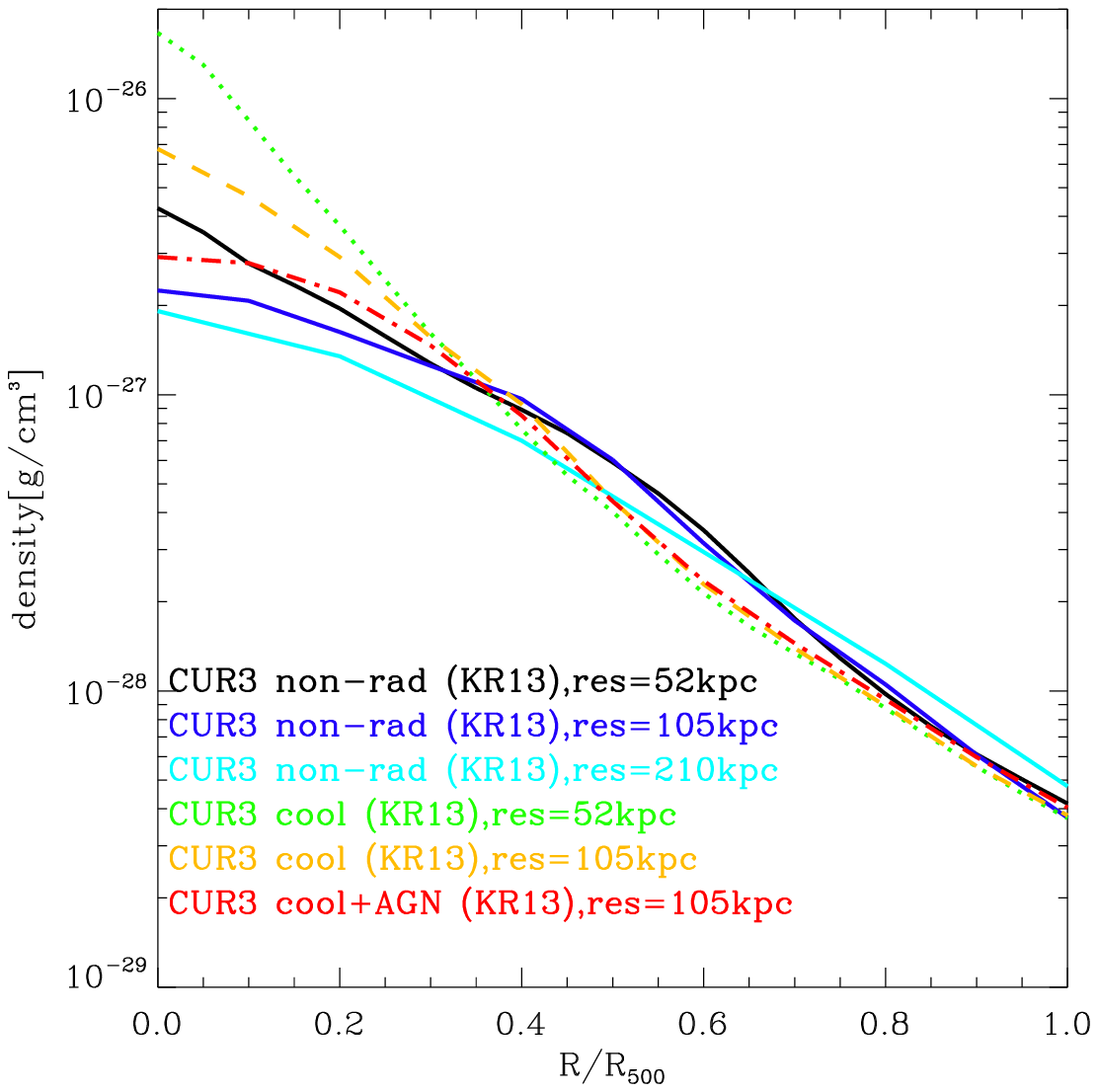}
      \includegraphics[width=0.24\textwidth]{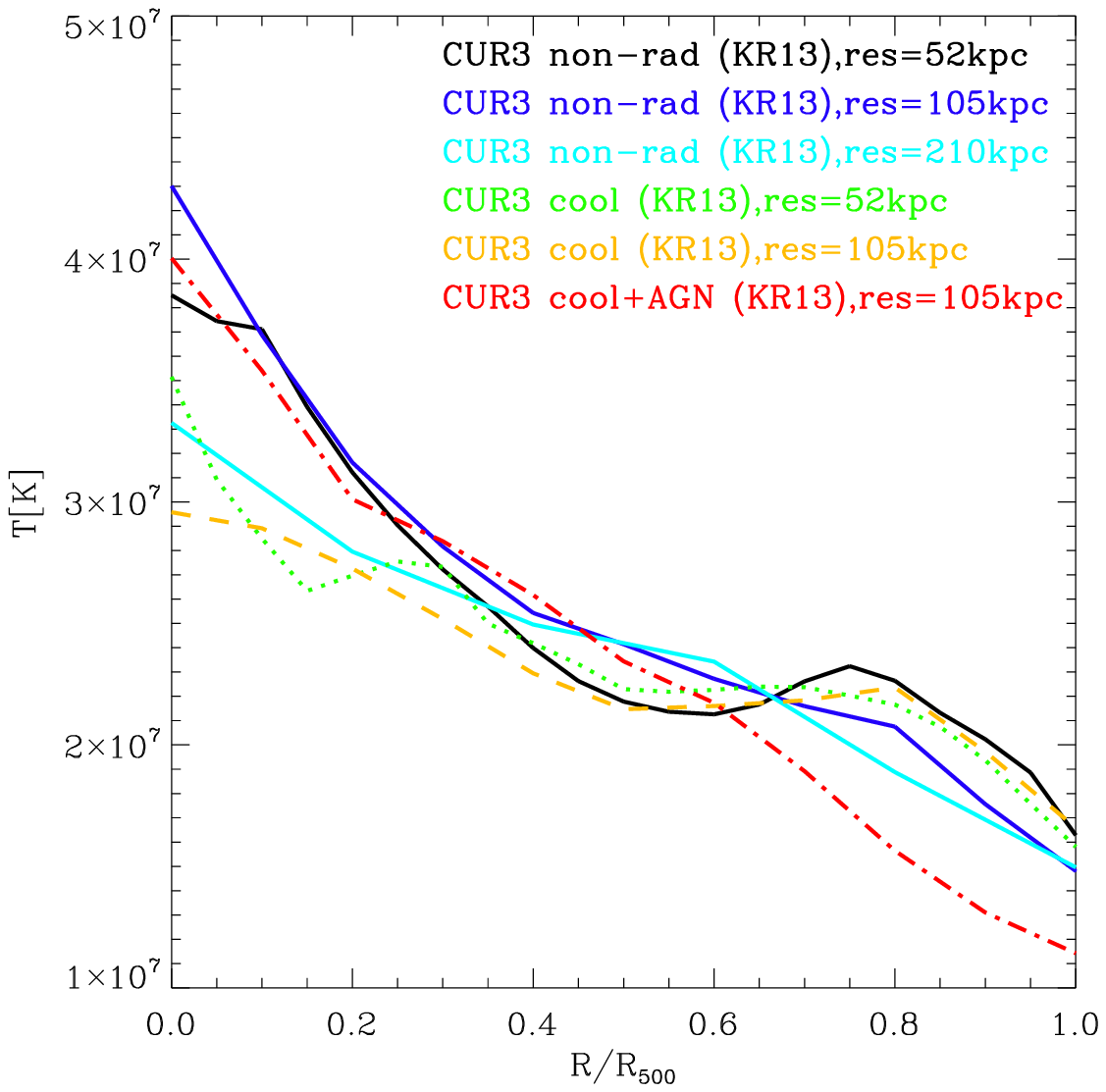}
      \includegraphics[width=0.24\textwidth]{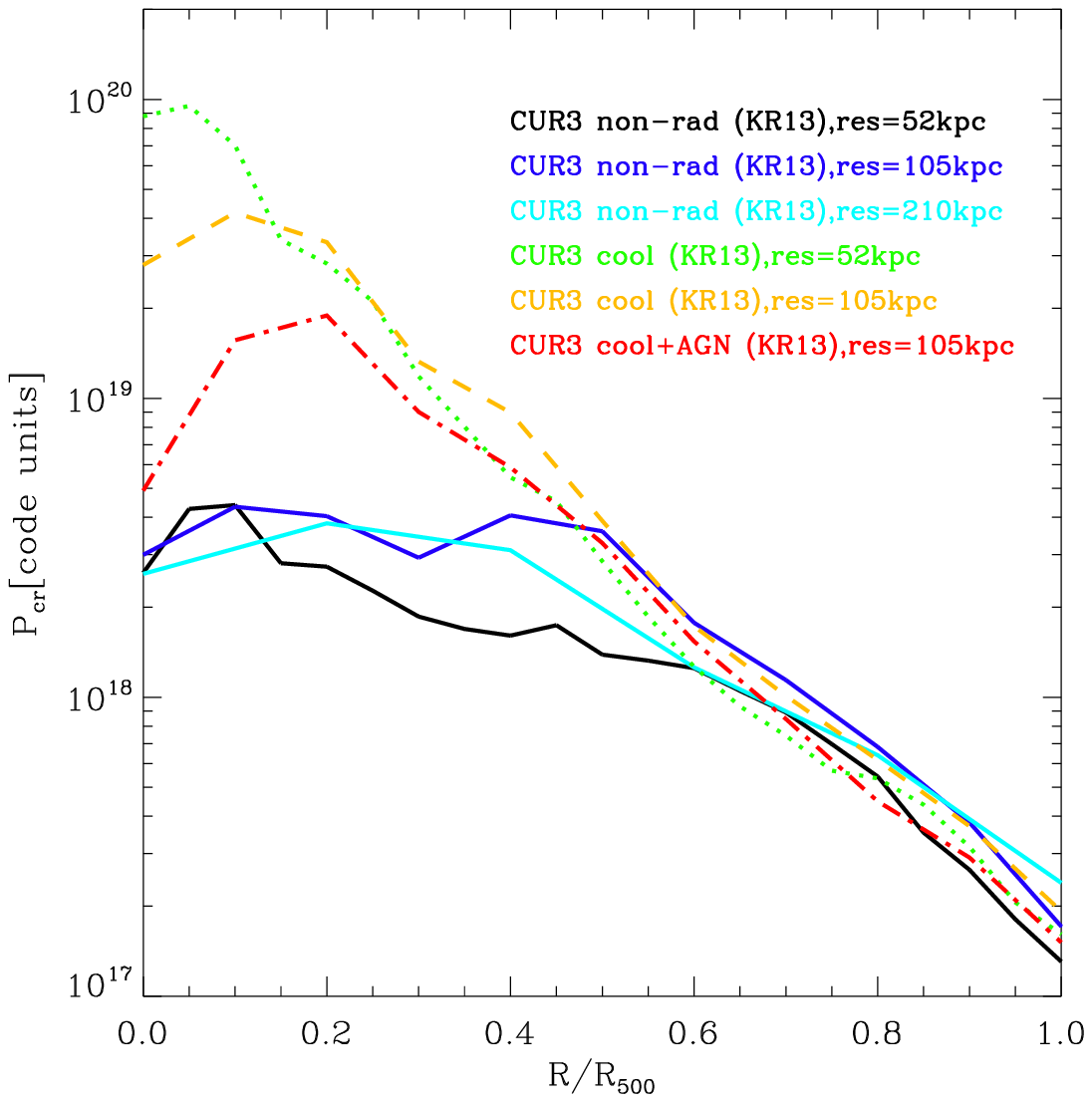}
           \includegraphics[width=0.24\textwidth]{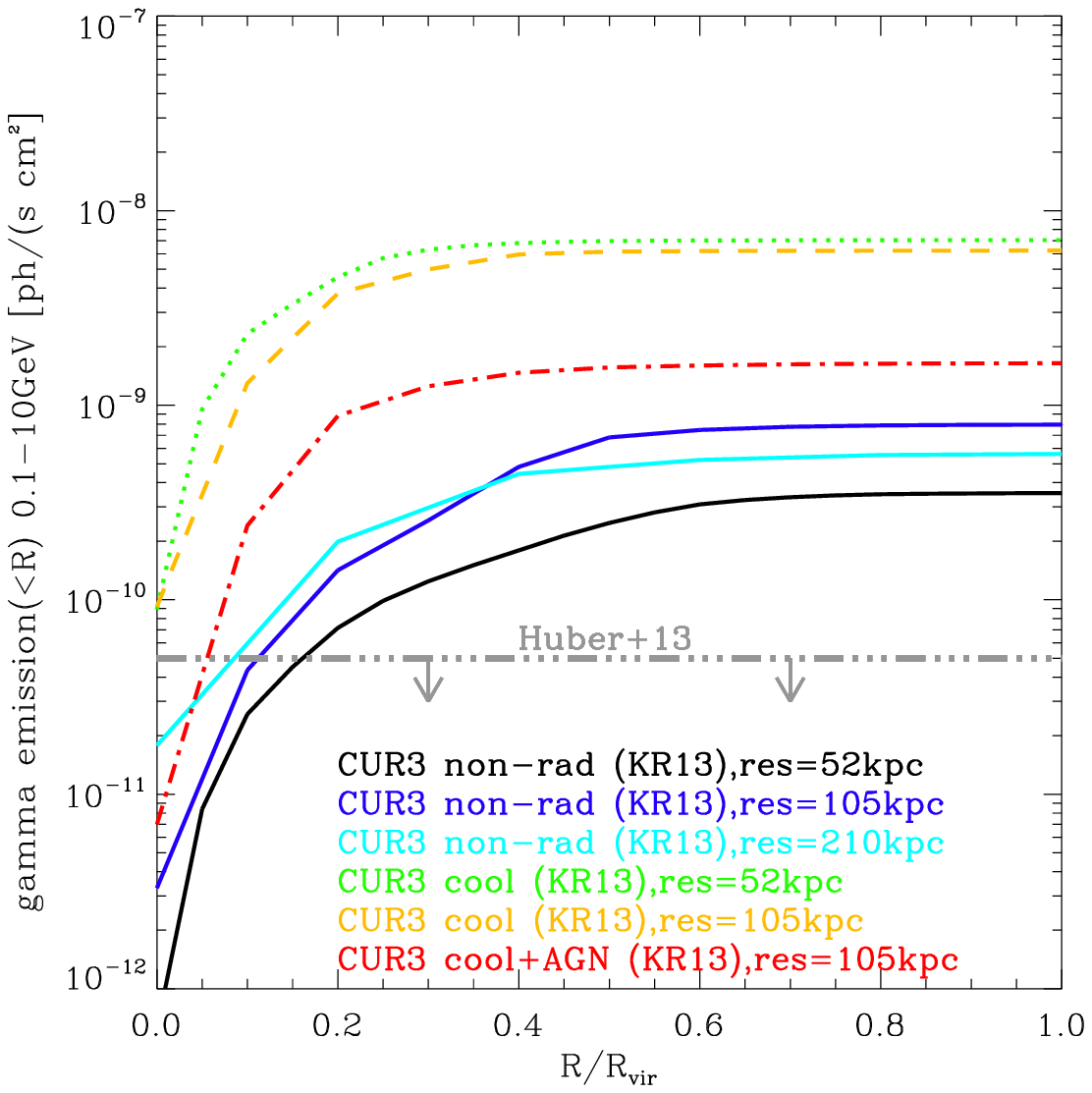}
    \caption{Radial profiles of gas density, gas temperature, CR-pressure and $\gamma$-ray emission inside the radius  for a $\sim 2 \times 10^{14} M_{\odot}$ simulated cluster at $z=0$, for different resolutions and physical prescriptions for baryons (see text). The additional horizontal line in the last panel marks the 
upper limit on the $\gamma$-ray emission for the stacking of clusters obtained by \citet{2013A&A...560A..64H}.}
  \label{fig:resolution}
\end{figure}

We use the smallest simulated volume in our suite of simulation (CUR3, with side $75 \rm Mpc$, see \citealt{scienzo14}) to compare the effects of resolutions and different physical prescriptions for baryon physics onto the radial distribution of baryons and CRs. 
Figure \ref{fig:resolution} shows the radial profiles of gas density, gas temperature, CR-pressure and $\gamma$-ray emission for the most massive halo formed in this box, a cluster with a total mass of  $\sim 2 \times 10^{14} M_{\odot}$ in a fairly relaxed dynamical state at $z=0$. We compare here two sets of data: a) non-radiative runs using the acceleration efficiency of \citet{kr13} for CRs, and increasing spatial resolution from
$210$ to $52.5$ kpc (i.e. from the number of cells/DM particles in the full volume goes from $256^3$ to $1024^3$); b) runs including equilibrium gas cooling and high redshift AGN feedback, as the main article \citep[see][for details]{scienzo14}, at the resolution of $52.5$ and $105$ kpc.
The trends which are most relevant for the main findings of our article are  the effects of resolution/physics in the spatial distribution of gas and CRs in the innermost cluster regions. 
The most significant differences connected to resolution and gas physics are limited to the innermost $\leq 0.3 R_{\rm 500}$ region, where the gas density is increased at most by $\sim 10$ times when cooling is not contrasted by AGN feedback, which also appears as a drop in the gas temperature in these runs. The outcome in the CR-pressure is also limited 
to the innermost region, where radiative runs without AGN feedback present a strong peak in the core, which
also corresponds to a $\sim 10$ times larger total $\gamma$-ray emission. When AGN feedback is applied, the $\gamma$-ray emission is only a factor $\sim 2$ above the prediction from non-radiative runs at the same resolution. 
The last panel of Fig. A1 shows indeed that the hadronic emission is mostly produced at $\geq 0.3-0.4 R_{\rm 500}$, where the differences played by resoluton and gas physics are small. 
In all cases, the predicted hadronic emission for this cluster is larger than the upper limits set by FERMI \citep[e.g.][]{2013A&A...560A..64H}.  These tests confirm that the problems in explaining the unseen population of CRs in galaxy clusters, outlined in our main article, are robust against variations in resolution and prescriptions for gas physics.

\section{Impact of CR physics on cluster scaling relations}
\label{appendix2}
\begin{figure}
    \includegraphics[width=0.45\textwidth]{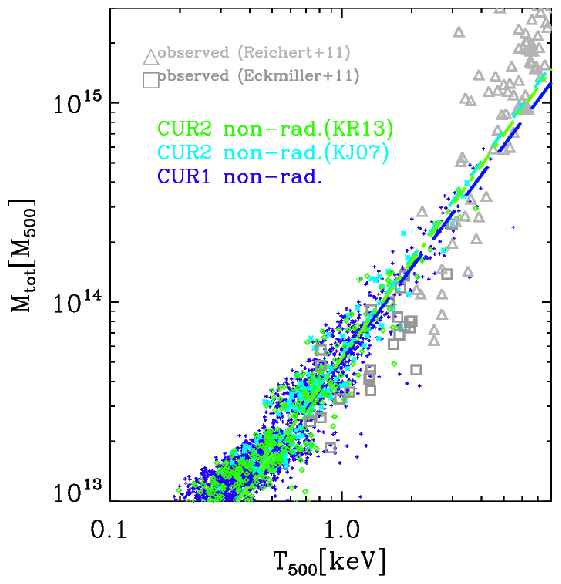}
    \caption{Mass-temperature scaling relation for the halos in the CUR1 and CUR2 volumes at $z=0$, where the effect of the \citet{kj07} and \citet{kr13} acceleration model for CRs are compared. The additional lines show the best fit of the simulated data, while the two set of gray symbols are for real cluster observations using CHANDRA by \citet{2011A&A...535A.105E} and \citet{2011A&A...535A...4R}. To better compare with the simulated cluster and minimise the effect of cosmic evolution, we only consider observed cluster in the $0 \leq z \leq 0.2 $ redshift range.}
  \label{fig:scaling_appendix}
\end{figure}

Figure \ref{fig:scaling_appendix} shows the $(M,T)$ scaling relation within $R_{\rm 500}$ for all halos in the non-radiatie CUR1 run ($300^3 \rm Mpc^3$) and in the non-radiative CUR2 runs ($150^3 \rm Mpc^3$) volume, where the impact on the different acceleration models \citep[][]{kj07,kr13} are compared.
For the typical pressure ratio between CRs and gas of these runs ($\leq 10\%$) the  dynamical impact of CRs is not enough to cause any significant departures
from the self-similar scaling in the $M \propto T^{3/2}$, nor to affect the mass-function of halos (not shown). In addition, the impact of CRs on the global scaling relation is the same also when more massive clusters (as in the CUR1 run) are included.

\end{document}